\documentclass{ws-ijmpd}
\usepackage{epsfig}
\usepackage{epstopdf}
\usepackage{color}

\begin{document}

\title{Anisotropic compact star with a linear pressure-density relationship}

\author{Shyam Das}
\address{Department of Physics, Malda College, Malda 732101, West Bengal, India\\shyam\_das@associates.iucaa.in}

\author{Bikram Keshari Parida}
\address{Department of Physics, Pondicherry University, Kalapet, Puducherry 605014, India\\parida.bikram90.bkp@gmail.com}   

\author{Koushik Chakraborty}
\address{Department of Physics, Government College of Education, Burdwan 713102, West Bengal, India\\koushik@associates.iucaa.in} 

\author{Saibal Ray}
\address{Department of Physics, Government College of Engineering and Ceramic Technology, Kolkata 700010, West Bengal, India \\saibal@associates.iucaa.in}

\maketitle	

\abstract{We present a model of compact astrophysical object under General Theory of Relativity using the anisotropic extension of Tolman IV solution. The anisotropy function, derived from the model, remains well behaved throughout the interior of the star. The model satisfies several necessary conditions for a physically realistic compact star. Physical viability of the model is verified specifically by plugging in the estimated parameter values of the Low Mass X-ray Binary (LMXB) candidate $4U~1608-52$. Our stability analysis of this star, by using various criteria for stability, provide satisfactory results. In connection to anisotropy, we compute the Tidal Love Number (TLN) for the compact stellar model and compare the calculated values with existing literature. }


\section{Introduction}
\label{sec:1}
~~~~~Study of static spherically symmetric perfect fluid sphere under General Theory of Relativity is an extensively explored field of research  for over a century. There are some comprehensive reviews of the important results regarding this field of research~\cite{Kramer,Delgaty}. In their review, Delgaty and Lake~\cite{Delgaty} pointed out six essential conditions for a physically realistic solution. Tolman's seminal solution \cite{TolmanIV}, commonly known as Tolman IV solution, is reported to satisfy all the six conditions of physical acceptibility.

This solution is being explored by some researchers in recent past. The Braneworld version of Tolman IV solution was presented by Ovalle and Linares \cite{Ovalle}. Singh et al.~\cite{Singh} investigated the behavior of the Tolman IV solution in bimetric gravity describing compact fluid sphere. Tolman obtained the solution for spherically symmetric static perfect fluid sphere taking into consideration isotropic principal pressure components. In the present paper, we like to obtain an anisotropic extension of Tolman's solution where anisotropy implies unequal principal stresses. In an interesting article, Grenon et al.~\cite{Grenon2008} derived a class of solutions which can be regarded as the generalization of Tolman IV solution, but can not be written in terms of isotropic coordinates. Recently, Sharif and Ama-Tul-Mughani \cite{Sharif2020} obtained anisotropic extension of Tolman IV solution using extended gravitational decoupling. Malaver~\cite{Malaver} presented a relativistic model of anisotropic quark star using Tolman IV like gravitational potential. Arias et al. \cite{Arias2020} used the principle of Gravitational Decoupling in the framework of the welknown approach of Minimal Geometric Deformation to obatin an anisotropic extension of Tolman IV solution. They reported extra packing of mass within the compact object.

There is a large body of literature exploring the anisotropic compact stars under General Relativity~\cite{Gleiser2003,Sharma7,Varela,Rahaman1,Maharaj,Karmakar,Ivanov2002,Mak1,Mak2}. Though the radial and transverse components of pressure are taken to be unequal in these models, spherical symmetry of the stars dictates the transverse components to be equal~\cite{Gleiser2002}. Herrera and Santos~\cite{Herrera97} reviewed the probable causes for the origin of local anisotropy inside compact objects. Several speculations are there, regarding the origin of anisotropy inside a compact star. Exotic phase transition at ultrahigh density in the core of the compact stars may lead to anisotropy~\cite{Sokolov}. Jones~\cite{Jones} predicted the presence of type II superconductor inside the compact stars leading to the anisotropy of stress tensor. Pion condensation may lead to the softening of the equation of state along the radial direction~\cite{Sawyer}. Type $3A$ superfluid~\cite{Kippen} might also be the possible origins of anisotropy. Ruderman~\cite{Ruderman} indicated that local anisotropy may develop in compact stars due to relativistic interaction between the nucleons in ultra dense matter inside the star. The neutron stars may possess magnetic field of the order of $10^{12} - 10^{13}$ G. There are examples like $SGR1806 - 20$ having estimated magnetic field $ > 10^{14}$ G~\cite{Kouveliotou98}.  Weber~\cite{Weber} predicted that inside the compact star anisotropy may arise due to strong magnetic field. Finally, it may be noted that scalar field in a Boson star may give rise to anisotropy~\cite{Liebling12}.

The Tidal Love Numbers (TLNs) illustrate the deformability of a compact star due to an external field which may be the gravitational field of a companion boundary. TLNs play a significant role in gravitational wave astronomy~\cite{Biswas19}. Flanagan et al.~\cite{Flanagan08} showed that these quantities provide significant information for constraining the equation of state of the compact stars. The nature of relativistic compact objects can be understood through their TLN values. In the limit of nonrotating black holes the TLNs are all zero~\cite{Binnington09,Damour09}. Cardoso et al.~\cite{Cardoso17} calculated the TLNs for various exotic compact objects like boson stars, gravastars, wormholes etc and black holes as well. Sennett et al.~\cite{Sennett17} computed tidal deformabilities of boson star, neutron star and black hole, thereby, indicating a novel method of distinguishing the three types of objects. There are other studies reporting that TLNs are zero for black holes, but have small finite value the exotic compact objects \cite{Maselli18,Pani17}. However, for the interested authors a detailed discussions are available in refs.~\cite{Yagi2017,Rahmansyah2020} in connection to the effects of the tidal forces and their impact on EOS. 

Here, we put forward a physically viable model of compact star, considering anisotropic pressures inside the star. With matter distribution following linear equation of state, we show that all the criteria for physical acceptability, proposed by Delgati and Lake~\cite{Delgaty} holds good for the proposed model. The interior matter distribution satisfies Null, Weak and Strong energy conditions of general relativity.  In particular, values of various physical quantities computed from the model by plugging in the estimated values of parameters for $4U~1608 - 52$~\cite{Guver10,Poutanen14}, are found to be complacent with the existing literature. We compute the TLNs for the model compact star and its computed values and plots, are found to agree with the existing predictions \cite{Cardoso17,Sennett17,Yazadjiev2018,Thiru18}.

The paper is organized as follows. In Section \ref{sec:2} the Einstein field equations describing a spherically symmetric static anisotropic matter distribution is given and thereafter by assuming a particular geometry and a linear equation of state (EOS), we have solved the system to generate a new model along with the related matching conditions. In Section \ref{sec:3} bounds on the physical parameters are sought for whereas the physical viability and stability analysis of our model have been studied in Section \ref{sec:4}. To understand the role of anisotropy we have investigated and calculated tidal Love number in Section \ref{sec:5} . Finally some conclusions have been made in discussion Section \ref{sec:6}.

\section{Einstein field equations and their solutions} \label{sec:2}
~~~~~We write the line element describing the interior space-time of a spherically symmetric star in standard coordinates $x^0 = t$,  $x^1=r$,  $x^2 = \theta$,  $x^3 = \phi$ as
\begin{equation}
ds^2 = -e^{\nu(r)}(r)dt^2 + e^{\lambda(r)}dr^2 + r^2(d\theta^2 + \sin^2{\theta}d\phi^2),\label{metric}
\end{equation}
where, $e^{\nu(r)}$ and  $e^{\lambda(r)}$ are the gravitational potential are yet to be determined. 

We assume that the matter distribution of the stellar interior is anisotropic in nature and described by an energy-momentum tensor of the form 
\begin{equation}
T_{ij} = (\rho + p_t)u_{i} u_{j} + p_{t} g_{i j} + (p_r - p_t)\chi_{i} \chi_{j},\label{eq2}
\end{equation}
where $\rho$ represents the energy-density, $p_r$ and $p_t$, respectively denote fluid pressures along the radial and transverse directions, $u^i$ is the $4$-velocity of the fluid and $\chi^i$ is a unit space-like $4$-vector along the radial direction so that $u^i u_i = -1$, $\chi^i \chi_j =1$ and $u^i\chi_j=0 $.

The Einstein field equations for the line element (\ref{metric}) are obtained as (in system of units having $ G =c=1$)
\begin{eqnarray}\label{g3}
8\pi \rho &=& \frac{\left(1 - e^{-\lambda}\right)}{r^2} + \frac{\lambda^{\prime}e^{-\lambda}}{r},\,\label{g3a} \\ \nonumber \\
8\pi p_r &=&  \frac{\nu^{\prime}e^{-\lambda}}{r} - \frac{\left(1 - e^{-\lambda}\right)}{r^2},\, \label{g3b}  \\  \nonumber \\
8\pi p_t &=& \frac{e^{-\lambda}}{4}\left(2\nu^{\prime\prime} + {\nu^{\prime}}^2  - \nu^{\prime}\lambda^{\prime} + \frac{2\nu^\prime}{r} - \frac{2\lambda^\prime}{r}\right), \label{g3c}
\end{eqnarray}
where primes $(')$ represent differentiation with respect to the radial coordinate $r$.

Making use of Eqs.~(\ref{g3b}) and (\ref{g3c}), we define the anisotropic parameter of the stellar system as
\begin{equation}
8\pi \Delta(r) = 8\pi (p_t-p_r) = \frac{e^{-\lambda}}{4} \left(2\nu^{\prime\prime} + {\nu^{\prime}}^2  - \nu^{\prime}\lambda^{\prime} - \frac{2}{r}(\nu'+\lambda') + \frac{4}{r^2}(e^{\lambda}-1)\right).\label{eq6} 
\end{equation}

The anisotropic force which is defined as $\frac{2 \Delta}{r}$ will be repulsive or attractive in nature depending upon whether $p_t>p_r$ or $p_t<p_r$.

Thus we have a system of four equations Eq.~(\ref{g3a})-Eq.~(\ref{eq6}) with $6$ independent variables, namely $e^{\lambda}$, $e^{\nu}$, $\rho$, $p_r$, $p_t$ and $\Delta$. We need to specify two of them to solve the system. In this  model we solve the system by assuming a particular metric anasatz $g_{rr}$ and the interior matter distribution to follow a linear equation of state.

Now, to develop a physically reasonable model of the stellar configuration, we assume that the metric potential $g_{rr}$ is given by
\begin{equation}
e^{\lambda(r)} = \frac{2 a \text{C} r^2+1}{\left(a \text{C} r^2+1\right) \left(1-B \text{C} r^2\right)-\text{C} \delta  r^2},\label{etlambda}
\end{equation}
where $a$, $C$ and $B$ are the constants to be determined from the matching conditions wheres $\delta$ is the anisotropic parameter. 

This particular  metric~\cite{Thiru18} is the anisotropic extension of the well known Tolman IV solution used to model realistic compact stellar object. In addition, to develop a stellar model we have prescribed a linear equation of state of the form
\begin{equation}
p_r = \alpha \rho +\beta,  \label{eos}
\end{equation}
where $\alpha$ and $\beta$  are constants. The idea behind this prescription is that the $\beta$ may act as a tunning parameter with a realistic physical bound so that the presented model is expected to provide wide range of possibilities. One can note that in the absence of $\beta$ the usual EOS can be recovered. Here $\alpha$ also free to assume in a definite form connecting to normal matter, stiff matter or exotic matter as required.  

Substituting Eq.~(\ref{etlambda}) in Eq.~(\ref{g3a}) and using Eqs.~(\ref{eos}) and~(\ref{g3b}), we have 
\begin{align}
\nu' &=-\dfrac{1}{8 \pi  \left(2 a \text{C} r^2+1\right) \left(\text{C} r^2 \left(a \left(B \text{C} r^2-1\right)+B+\delta \right)-1\right)} \left[ 8 \pi  r \left(2 a \text{C} r^2+1\right)\right.\nonumber\\
&\times \left(\text{C} \left(a r^2 (2 \beta +B \text{C})+a+B+\delta \right)+\beta\right) \alpha  +\text{C} r \left(a \left(\text{C} r^2 \left(a \left(6 B \text{C} r^2+2\right)+7 B+2 \delta \right)\right.\right.\nonumber\\
&\left.\left.\left.+3\right)+3 (B+\delta )\right) \right].\label{soln}
\end{align}

Integrating we have
\begin{align}
\nu &=C_{2}-\frac{1}{32 \pi  B \text{C} \sqrt{a^2+2 a (B-\delta )+(B+\delta )^2}} \left[ -4 \alpha  B \text{C} \sqrt{a^2+2 a (B-\delta )+(B+\delta )^2}\right.\nonumber\\
&\times \log \left(2 a \text{C} r^2+1\right)+ \left[ a (16 \pi  \beta +3 (\alpha +8 \pi ) B \text{C})+(\alpha +8 \pi ) B^2 \text{C}\right.\nonumber\\
& +16 \pi  \beta  \left(\sqrt{2 \delta  (B-a)+(a+B)^2+\delta ^2}-\delta \right)+B \text{C} \left(5 \alpha  \sqrt{2 \delta  (B-a)+(a+B)^2+\delta ^2}\right.\nonumber\\
&\left.\left.+8 \pi  \sqrt{2 \delta  (B-a)+(a+B)^2+\delta ^2}+(\alpha +8 \pi ) \delta \right)\right]\log \left(a \left(2 B \text{C} r^2-1\right)\right.\nonumber\\
&\left.-\sqrt{2 \delta  (B-a)+(a+B)^2+\delta ^2}+B+\delta \right)+ \left[16 \pi  \beta  (\delta -a)-(\alpha +8 \pi ) B^2 \text{C} \right.\nonumber\\
&+8 \pi  B \text{C} \sqrt{a^2+2 a (B-\delta )+(B+\delta )^2}-(\alpha +8 \pi ) B \text{C} (3 a+\delta )\nonumber\\
&\left.+16 \pi  \beta  \sqrt{a^2+2 a (B-\delta )+(B+\delta )^2}++5 \alpha  B \text{C} \sqrt{a^2+2 a (B-\delta )+(B+\delta )^2}\right]\nonumber\\
&\left.\log \left(a \left(2 B \text{C} r^2-1\right)+\sqrt{2 \delta  (B-a)+(a+B)^2+\delta ^2}+B+\delta \right)\right],\label{soln1}
\end{align}
and hence
\begin{align}
e^{\nu(r)} &=e^{C_{2}} \left(2 a \text{C} r^2+1\right)^{\text{$\eta $1}} \left(-\sqrt{a^2+2 a (B-\delta )+(B+\delta )^2}+a \left(2 B \text{C} r^2-1\right)+B+\delta \right)^{-\text{$\eta $2}} \nonumber\\
&\times\left(\sqrt{a^2+2 a (B-\delta )+(B+\delta )^2}+a \left(2 B \text{C} r^2-1\right)+B+\delta \right)^{\text{$\eta $3}},\label{soln}
\end{align}
where\\ 
$\text{$\eta $1}=\frac{4 \alpha  B \text{C} \sqrt{a^2+2 a (B-\delta )+(B+\delta )^2}}{\xi }$,\\
$\text{$\eta $2}=\dfrac{1}{\xi}\left[16 \pi  a \beta +16 \pi  \beta  \left(\sqrt{2 \delta  (B-a)+(a+B)^2+\delta ^2}-\delta \right)\right.\\
+B \text{C} \left(\alpha  \left(5 \sqrt{2 \delta  (B-a)+(a+B)^2+\delta ^2}+\delta \right)+8 \pi  \left(\sqrt{2 \delta  (B-a)+(a+B)^2+\delta ^2}+\delta \right)\right)\\
\left.+3 a (\alpha +8 \pi ) B \text{C}+(\alpha +8 \pi ) B^2 \text{C}\right]$,\\
$\text{$\eta $3}=\\ \dfrac{1}{\xi} \left[-16 \pi  \beta  \left(\sqrt{a^2+2 a (B-\delta)+(B+\delta )^2}+\delta \right)+16 \pi  a \beta +3 a (\alpha +8 \pi ) B \text{C}+(\alpha +8 \pi ) B^2 \text{C}\right.\\
\left.+B \text{C} \left(-5 \alpha  \sqrt{a^2+2 a (B-\delta )+(B+\delta )^2}-8 \pi  \sqrt{a^2+2 a (B-\delta )+(B+\delta )^2}+\alpha  \delta +8 \pi  \delta \right)\right]$,
$\xi =32 \pi  B \text{C} \sqrt{a^2+2 a (B-\delta )+(B+\delta )^2}$,\\
and  \\
$A=e^{C2}$ is a constant of integration. 

Interestingly, here the constant $A$ does not appear in the expressions of physical parameters, e.g., $\rho$, $p_r$, $p_t$, $\Delta$, $m(r)$, $\frac{dp_r}{d\rho}$, $\frac{dp_t}{d\rho}$. 

Consequently, the physical quantities are obtained as
\begin{align}
\rho &=\frac{\text{C} \left(a \left(\text{C} r^2 \left(a \left(6 B \text{C} r^2+2\right)+7 B+2 \delta \right)+3\right)+3 (B+\delta )\right)}{8 \pi \left(2 a \text{C} r^2+1\right)^2},\label{den}\\
p_r &=\frac{\alpha  \text{C} \left(a \left(\text{C} r^2 \left(a \left(6 B \text{C} r^2+2\right)+7 B+2 \delta \right)+3\right)+3 (B+\delta )\right)}{8 \pi  \left(2 a \text{C} r^2+1\right)^2}+\beta,\label{radpres}\\
p_t &=\dfrac{1}{2048 \left(2 \pi  a \text{C} r^2+\pi \right)^3 \left(\text{C} r^2 \left(a \left(B \text{C} r^2-1\right)+B+\delta \right)-1\right)}\nonumber\\
&\times\left[4 a^4 \text{C}^4 r^6 \xi_{1} -9 \alpha ^2 \text{C}^2 r^2 (B+\delta )^2-48 \pi  \alpha  \text{C} (B+\delta ) \left(\beta  r^2+2\right)\right.\nonumber\\
&\left.-64 \pi ^2 \left(4 \beta +r^2 \left(\beta ^2+3 \text{C}^2 (B+\delta )^2\right)\right)+ 4 a^3 \text{C}^3 r^4 \xi_{2}\right],\label{tangpres}\\	
\Delta &= (p_t - p_r),\label{ani}
\end{align}
where\\
$\xi_{1} = -16 \pi  \alpha  \left(r^2 \left(2 \beta +B \text{C} \left(6 \beta  r^2+11\right)\right)+1\right)-\left(\alpha +3 \alpha  B \text{C} r^2\right)^2\\
-64 \pi ^2 \left(r^2 \left(12 \beta +B \text{C} \left(3 B \text{C} r^2+4\right)+4 \beta ^2 r^2\right)+1\right)$,\\
$\xi_{2} =\\ -128 \pi ^2 \left(r^2 \left(4 B^2 \text{C}^2 r^2+B \text{C} \left(4-r^2 (\beta -2 \text{C} \delta )\right)+\text{C} \delta +2 \beta  \left(r^2 (2 \beta -\text{C} \delta )+7\right)\right)+1\right)\\
+\alpha ^2 \left(-\left(3 B \text{C} r^2+1\right)\right) \left(\text{C} r^2 (7 B+2 \delta )+3\right)+8 \pi  \alpha  \left(r^2 \left(-10 \beta +2 \text{C} \delta  \left(r^2 (B \text{C}-2 \beta )-2\right)\right.\right.\\
\left.\left.+B \text{C} \left(r^2 (B \text{C}-26 \beta )-56\right)\right)-3\right)$.\\

The parameter $\beta$ can be expressed as $\beta = -\alpha \rho_R $, where $R$ is the radius of the star and $\rho_R$ is the surface density given by
\begin{equation}
\rho_R=\frac{\text{C} \left(a \left(\text{C} R^2 \left(a \left(6 B \text{C} R^2+2\right)+7 B+2 \delta \right)+3\right)+3 (B+\delta )\right)}{8 \pi \left(2 a \text{C} R^2+1\right)^2}.\label{rhoR}
\end{equation}

This ensures that the radial pressure $p_r(r=R) = 0$. The central density $\rho(r=0)$ can be obtained from Eq.~(\ref{den}) as
\begin{equation}
\rho_c = \text{C} (3 a+3 (B+\delta ))/8 \pi. \label{cden}
\end{equation}
 
For $\delta=0$, i.e., for isotropic case the above condition reads as 
\begin{equation}
\rho_c = \text{C} (3 a+3 B)/8 \pi. \label{cden}
\end{equation}

In this connection, it is to note that the anisotropy vanishes at the centre, i.e., $ \Delta (r=0) = 0$.

The mass contained within a sphere of radius $r$ is defined as
\begin{equation}
\label{massfn} m(r)= \frac{1}{2} \int\limits_0^r\omega^2
\rho(\omega)d\omega,
\end{equation}
which on integration yields
\begin{equation}
m(r) = \frac{\text{C} r^3 \left(a B \text{C} r^2+a+B+\delta \right)}{16 \pi \left(2 a \text{C} r^2+1\right)},\label{mass}
\end{equation}
obviously, $m(r=0) = 0$.

At this juncture we need to match the interior solution to the Schwarzschild exterior
\begin{equation}
 ds^2 = - \left(1 - \frac{2M}{r}\right) dt^2 + \left(1 - \frac{2M}{r}\right)^{-1}dr^2 - r^2 (d\theta^2 + sin^2 \theta~d\phi^2),
\label{mm}
\end{equation}
across the boundary $R$ where $M = m(R) $ is the total mass.

The matching conditions determine the constants as 
\begin{equation}
\text{C}=\frac{\sqrt{R^4 \left((a (R-4 M)+R (B+\delta ))^2+8 a B M R\right)}+R^3 (-(a+B+\delta ))+4 a M R^2}{2 a B R^5},\label{capital_B}
\end{equation}

\begin{align}
 \beta &= \dfrac{\alpha}{8 \pi  a R^6 (2 a+B+2 \delta )} \left[ a^2 R^2 \left(16 M^2-12 M R+R^2\right) + R (B+\delta ) \chi_{1}\right.\nonumber\\
 &\left.+a (2 R^4 (B+\delta )-2 M R^3 (B+6 \delta )- \chi_{2} - \chi_{3})\right], \label{C}
\end{align}

\begin{eqnarray}
A &=& \left(1-\frac{2 M}{R}\right) \left(2 a \text{C} R^2+1\right)^{-\alpha } \left(-\sqrt{a^2+2 a (B-\delta )+(B+\delta )^2}+a \left(2 B \text{C} R^2-1\right)+B+\delta
   \right)^{\Omega_{1}}\nonumber\\
   && \times \left(\sqrt{a^2+2 a (B-\delta
   )+(B+\delta )^2}+a \left(2 B \text{C} R^2-1\right)+B+\delta \right)^{\Omega_{2}},
\end{eqnarray}
where\\
$\chi_{1} = R^3 (B+\delta )-\sqrt{R^4 \left(a^2 (R-4 M)^2+2 a R (R (B+\delta )-4 \delta  M)+R^2 (B+\delta )^2\right)}$,\\
$\chi_{2} = 4 M \sqrt{R^4 \left(a^2 (R-4 M)^2+2 a R (R (B+\delta )-4 \delta  M)+R^2 (B+\delta )^2\right)}$,\\
$\chi_{3} = R \sqrt{R^4 \left(a^2 (R-4 M)^2+2 a R (R (B+\delta )-4 \delta  M)+R^2 (B+\delta )^2\right)}$,\\
$\Omega_{1} =\\ \frac{2 \beta  \left(\sqrt{a^2+2 a (B-\delta )+(B+\delta )^2}-\delta \right)+B \text{C} \left((5 \alpha +1) \sqrt{a^2+2 a (B-\delta )+(B+\delta )^2}+(\alpha +1)
   \delta \right)+2 a \beta +3 a (\alpha +1) B \text{C}+(\alpha +1) B^2 \text{C}}{4 B \text{C} \sqrt{a^2+2 a (B-\delta )+(B+\delta )^2}}$,\\
   $\Omega_{2}=\\ \frac{2 \beta  \left(\sqrt{a^2+2 a (B-\delta )+(B+\delta )^2}+\delta \right)-B \text{C}
   \left((\alpha +1) \delta -(5 \alpha +1) \sqrt{a^2+2 a (B-\delta )+(B+\delta )^2}\right)-2 a \beta -3 a (\alpha +1) B \text{C}-(\alpha +1) B^2 \text{C}}{4 B \text{C} \sqrt{a^2+2 a (B-\delta )+(B+\delta )^2}}.$

\section{Bounds on the model parameters} \label{sec:3}
~~~~~For a physically acceptable stellar model, it is reasonable to assume that the following conditions should be satisfied~\cite{Delgaty}:
(i) $\rho > 0$, $p_r > 0$, $p_t > 0$; (ii) $\rho' < 0$, $p_r' < 0$, $p'_t < 0$; (iii) $ 0 \leq \frac{dp_r}{d\rho} \leq 1$; $ 0 \leq \frac{dp_t}{d\rho} \leq 1$ and (iv) $\rho+p_r+2p_t > 0$. In addition, it is expected that the solution should be regular and well-behaved at all interior points of the stellar configuration. Based on the above requirements, bounds on the model parameters are obtained in this section.
\begin{enumerate}

\item \textbf{Regularity Condition:}
 \begin{enumerate}
  \item The metric potentials $ e^{\lambda(r)} > 0 $, $ e^{\nu(r)} > 0 $ for $ 0 \leq r \leq R $. These features are depicted in Fig. 1.\\

%%%%%%%%%%%%%%%%%%%%%%%%%%%%%%%%%%%%%%%%%%%%%%%%%%
\begin{figure}[h]
\centering
\includegraphics[width=0.4\columnwidth]{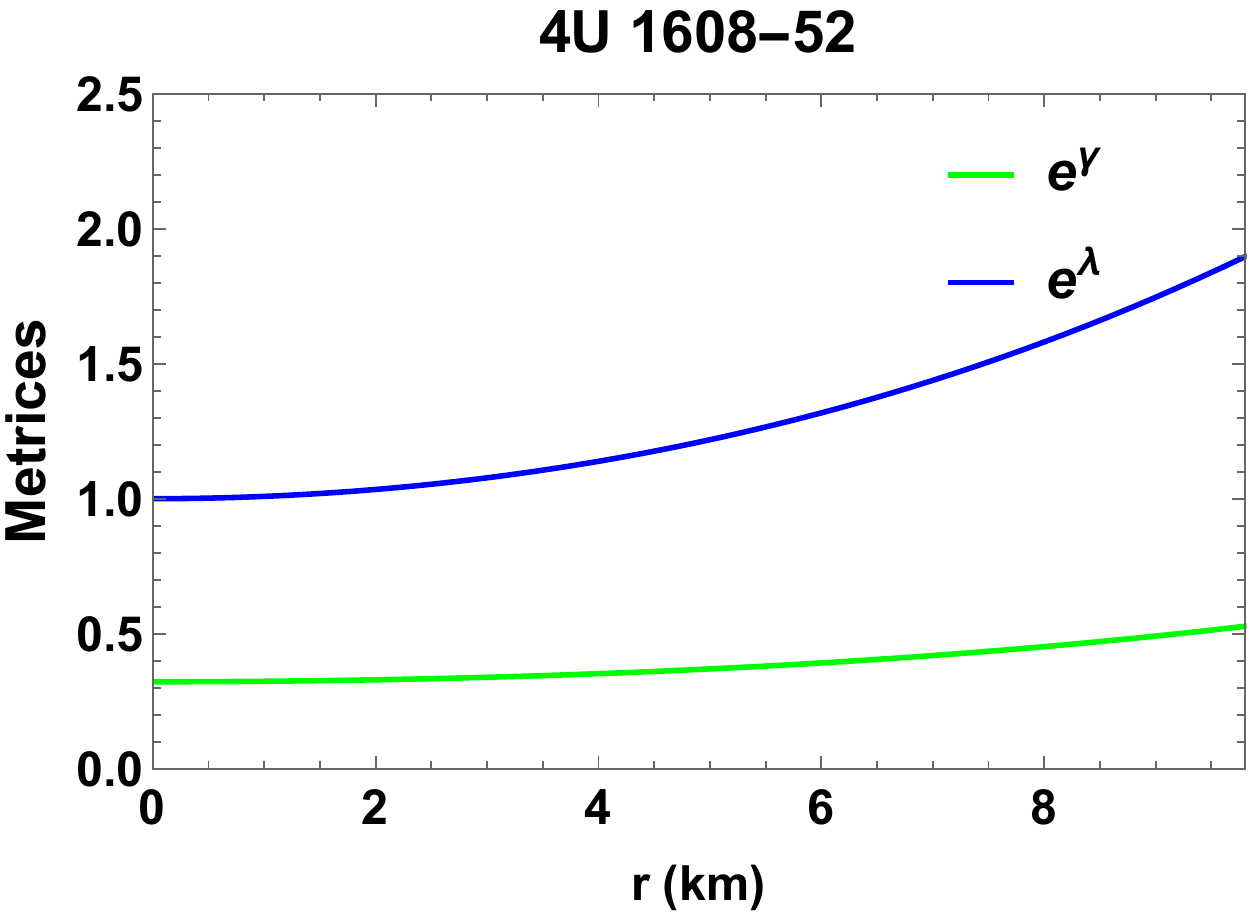}
\includegraphics[width=0.4\columnwidth]{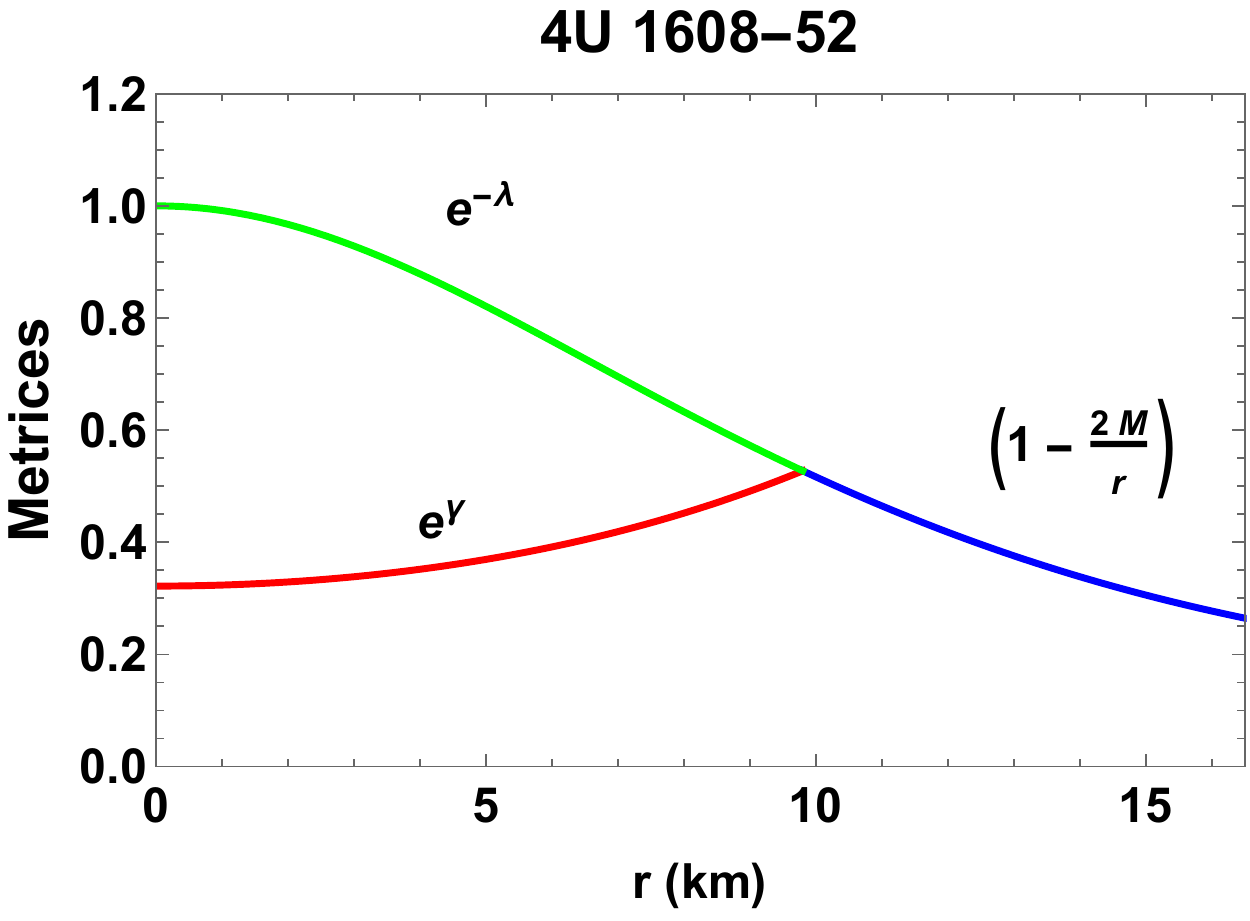}
\caption{The metric potentials $e^\nu$ and $e^\lambda$ are plotted against $r$ inside the stellar interior (left panel) and matching of the matrices at the boundary (right panel).}  \label{fig:metrices}
\end{figure}
%%%%%%%%%%%%%%%%%%%%%%%%%%%%%%%%%%%%%%%%%%%%%%%%%%

For appropriate choice of the model parameters, the above requirements are fulfilled in our model. The gravitational potentials in this model satisfy 
\begin{align}
e^{\nu(0)}&=A (-\sqrt{2 \delta  (B-a)+(a+B)^2+\delta ^2}-a+B+\delta)^{-\zeta_{1}}\nonumber\\
&\times(\sqrt{2 \delta  (B-a)+(a+B)^2+\delta ^2}-a+B+\delta)^{-\zeta_{2}},
	\end{align}
where\\
$\zeta_{1} = \dfrac{1}{\xi}\left[16 \pi  a \beta +16 \pi  \beta  \left(\sqrt{2 \delta  (B-a)+(a+B)^2+\delta ^2}-\delta \right)\right.\\
	+B \text{C} \left(\alpha  \left(5 \sqrt{2 \delta  (B-a)+(a+B)^2+\delta ^2}+\delta \right)\right.\\
\left.\left.	+8 \pi  \left(\sqrt{2 \delta  (B-a)+(a+B)^2+\delta ^2}+\delta \right)\right)+3 a (\alpha +8 \pi ) B \text{C}+(\alpha +8 \pi ) B^2 \text{C} \right],$
$\zeta_{2} = \dfrac{1}{\xi} \left[ 16 \pi  \beta  \sqrt{a^2+2 a (B-\delta )+(B+\delta )^2}+5 \alpha  B \text{C} \sqrt{a^2+2 a (B-\delta )+(B+\delta )^2}\right.\\
+8 \pi  B \text{C} \sqrt{a^2+2 a (B-\delta )+(B+\delta )^2}+16 \pi  \beta  (\delta -a)+(\alpha +8 \pi ) B (-\text{C}) (3 a+\delta )\\
\left.-(\alpha +8 \pi ) B^2 \text{C}\right]$,
which is a constant. \\

Again $e^{\lambda(0)}=1$, i.e., finite at the center ($r=0$) of the stellar configuration. Also one can easily check that $(e^{\nu(r)})'_{r=0}=(e^{\lambda(r)})'_{r=0}=0$. These imply that the metric is regular at the center and well behaved throughout the stellar interior which will be shown graphically.\\

 \item $ \rho (r) \geq 0, ~~ p_r (r) \geq 0, ~~ p_t (r) \geq 0 $ for $ 0 \leq r \leq R $.\\

From Eq.~(\ref{den}), we note that density remains positive if $a > 0$. Equation~(\ref{radpres}) shows that since $dp_r/d\rho (r=0)=\alpha$ is the sound speed must be between $0$ and $1$, so $0<\alpha <1$. From equation (\ref{tangpres}), we have
\begin{eqnarray}
p_t(r = 0) &=& \frac{3 \alpha  \text{C} (a+B+\delta )+8 \pi  \beta }{64 \pi ^2}. \label{tangpresatcentre}
\end{eqnarray}

We note that for $ $  the tangential pressure remain positive the centre $r=0$. Fulfillment of the requirements throughout the star can be shown by graphical representation.\\

\item $ p_r (r = R) = 0. $ \\

From Eq.~(\ref{radpres}), we note that the radial pressure vanishes at the boundary $R$ if we set $\beta =- \alpha \rho_R $, where $\rho_R $ is the surface density. Also at $p_r (r = 0) >0$  hence $\frac{\alpha  \text{C} (3 a+3 (B+\delta ))}{8 \pi }+\beta>0$
\end{enumerate}

For isotropic cases the above equation reduces to
$\frac{\alpha  \text{C} (3 a+3 B)}{8 \pi }+\beta>0$
\begin{eqnarray}
p_t(r = 0) &=& 3 \alpha  \text{C} (a+B)+8 \pi  \beta >0. \label{tangpresatcentre0}
\end{eqnarray}

\item \textbf{Causality Condition:}
The causality condition demands that $ 0 \leq \frac{dp_r}{d\rho} \leq 1$;~$ 0 \leq \frac{dp_t}{d\rho} \leq 1$ at all interior points of the star. Hence from Eq. (\ref{eos}) we have
\begin{equation}
\frac{dp_r}{d\rho}=\alpha.
\end{equation}

Similarly the expression for $\frac{dp_t}{d\rho}$ can also be calculated (See Appendix A for detailed calculation). 

 At the centre $r=0$, $\frac{dp_t}{d\rho}>0$, i.e.
\begin{align}
\frac{dp_t}{d\rho} &=-\dfrac{1}{20 a \text{C}^2 (2 a+B+2 \delta )}\left[a^2 \left(9 \alpha ^2-68 \alpha +3\right) \text{C}^2+a \text{C} \left(\left(18 \alpha ^2-16 \alpha +6\right) B \text{C}\right.\right.\nonumber\\
&\left.+2 \left((3 \alpha +2) \beta +\left(9 \alpha
   ^2-28 \alpha +3\right) \text{C} \delta \right)\right)+\beta ^2+3 \left(3 \alpha ^2+4 \alpha +1\right) B^2 \text{C}^2\nonumber\\
   &+2 B \text{C} \left((3 \alpha +2) \beta +3 \left(3
   \alpha ^2+4 \alpha +1\right) \text{C} \delta \right)+9 \alpha ^2 \text{C}^2 \delta ^2+12 \alpha  \text{C}^2 \delta ^2\nonumber\\
   &\left.+3 \text{C}^2 \delta ^2+6 \alpha  \beta  \text{C}
   \delta +4 \beta  \text{C} \delta \right]>0.
\end{align}

For isotropic case ($\delta=0$)
\begin{eqnarray}
\frac{dp_t}{d\rho} &=&\dfrac{- 1}{20 a \text{C}^2 (2 a+B)}\left[a^2 \left(9 \alpha ^2-68 \alpha +3\right) \text{C}^2+a \text{C} \left(2 (3 \alpha +2) \beta +\left(18 \alpha ^2-16 \alpha +6\right) B \text{C}\right)\right.\nonumber\\
&&\left.+\beta ^2+3
   \left(3 \alpha ^2+4 \alpha +1\right) B^2 \text{C}^2+2 (3 \alpha +2) \beta  B \text{C}\right]>0.
\end{eqnarray}

Also according to Zeldovich's condition~\cite{Zeldovich1,Zeldovich2}, $p_r/\rho$ must be $\leq 1$ at the center. Therefore, $\frac{3 a \alpha  \text{C}+8 \pi  \beta +3 \alpha  B \text{C}+3 \alpha  \text{C} \delta }{3 \text{C} (a+B+\delta )}\leq 1$. For $\delta=0$ $\frac{3 a \alpha  \text{C}+8 \pi  \beta +3 \alpha  B \text{C}}{3 a \text{C}+3 B \text{C}}\leq 1$.

\item \textbf{Energy Condition:}
For an anisotropic fluid sphere for being physically accepted matter composition, all the energy conditions, namely Weak Energy Condition (WEC), Null Energy Condition (NEC), Strong Energy Condition (SEC) and Dominant Energy Condition (DEC) are satisfied if and only if the following inequalities hold simultaneously in every point inside the fluid sphere.\\

(1) NEC : $\rho + p_r \geq  0$; $\rho + p_t \geq  0$, \\
(2) WEC : $p_r + \rho >  0, \rho > 0$, \\
(3) SEC : $\rho + p_r  \geq 0 , \rho + p_r + 2 p_t \geq 0$, \\
(4) DEC : $\rho > |p_r|,~\rho > |p_t|$.\\

%%%%%%%%%%%%%%%%%%%%%%%%%%%%%%%%%%%%%%%%%%%%%%%%%%%%%%
\begin{figure}[h]
\centering
\includegraphics[width=0.4\columnwidth]{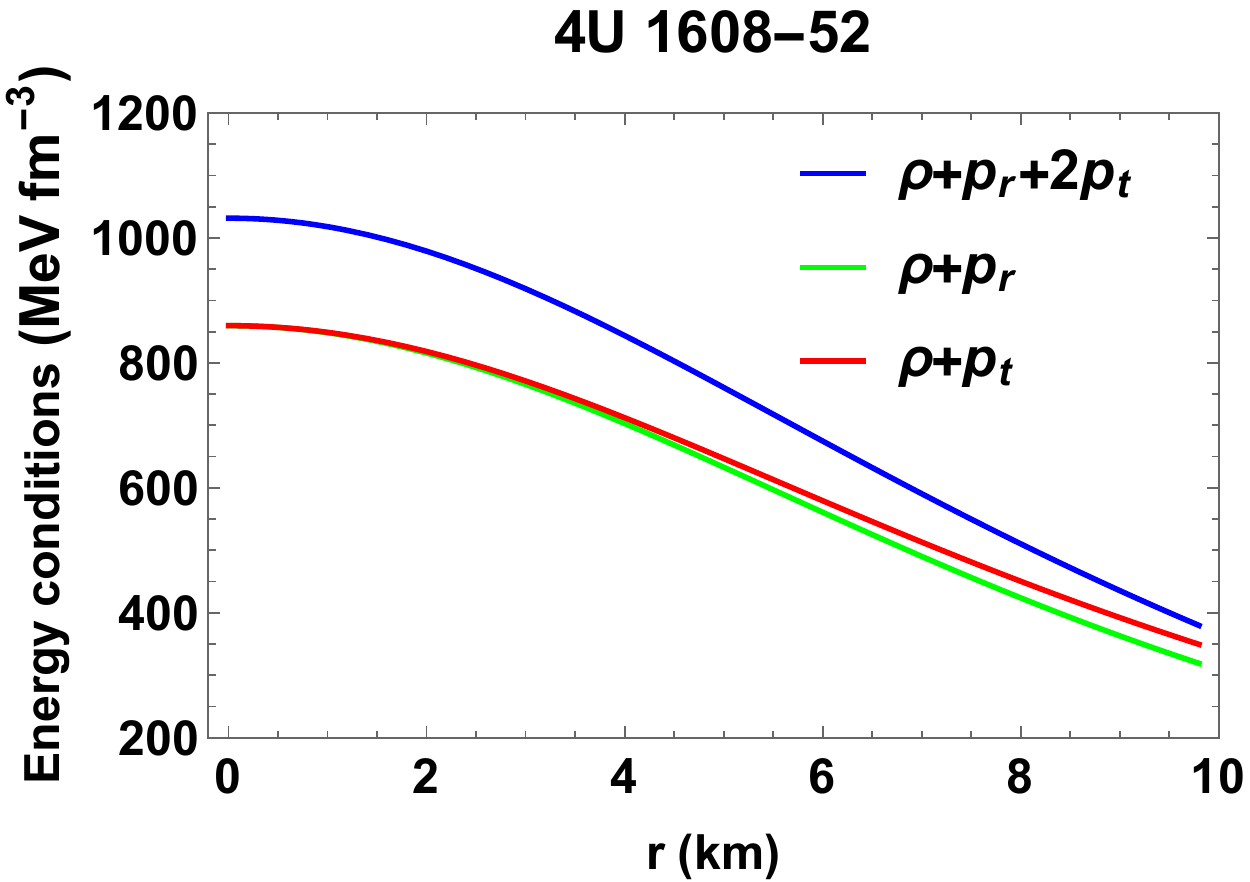}
\caption{Verification of the energy conditions w.r.t. the radial coordinate $r$.} \label{fig:energycondition_figure}
\end{figure}
%%%%%%%%%%%%%%%%%%%%%%%%%%%%%%%%%%%%%%%%%%%%%%%%%%%%%%

We have from SEC
\begin{eqnarray}
\rho + p_r + 2 p_t (r = 0) &=& 3 a (\alpha +4 \pi  (\alpha +1)) \text{C}+8 \pi  (1+4 \pi ) \beta\nonumber\\
&&+3 (\alpha +4 \pi  (\alpha +1)) B \text{C}+3 (\alpha +4 \pi  (\alpha +1)) \text{C} \delta>0,
\label{strngenergcondatctr}
\end{eqnarray}

We have shown energy conditions in Fig.~\ref{fig:energycondition_figure} for the compact stars $4U~1608-52$. \\

\item \textbf{Monotony condition:}

A realistic stellar model should have the following properties:

 $ \frac{d\rho}{dr} \leq 0,~\frac{dp_r}{dr} \leq 0,~\frac{dp_t}{dr} \leq 0$ for $ 0 \leq r \leq R $.\\

%%%%%%%%%%%%%%%%%%%%%%%%%%%%%%%%%%%%%%%%%%%%%%%%%%
\begin{figure}[h]
\centering
\includegraphics[width=0.4\columnwidth]{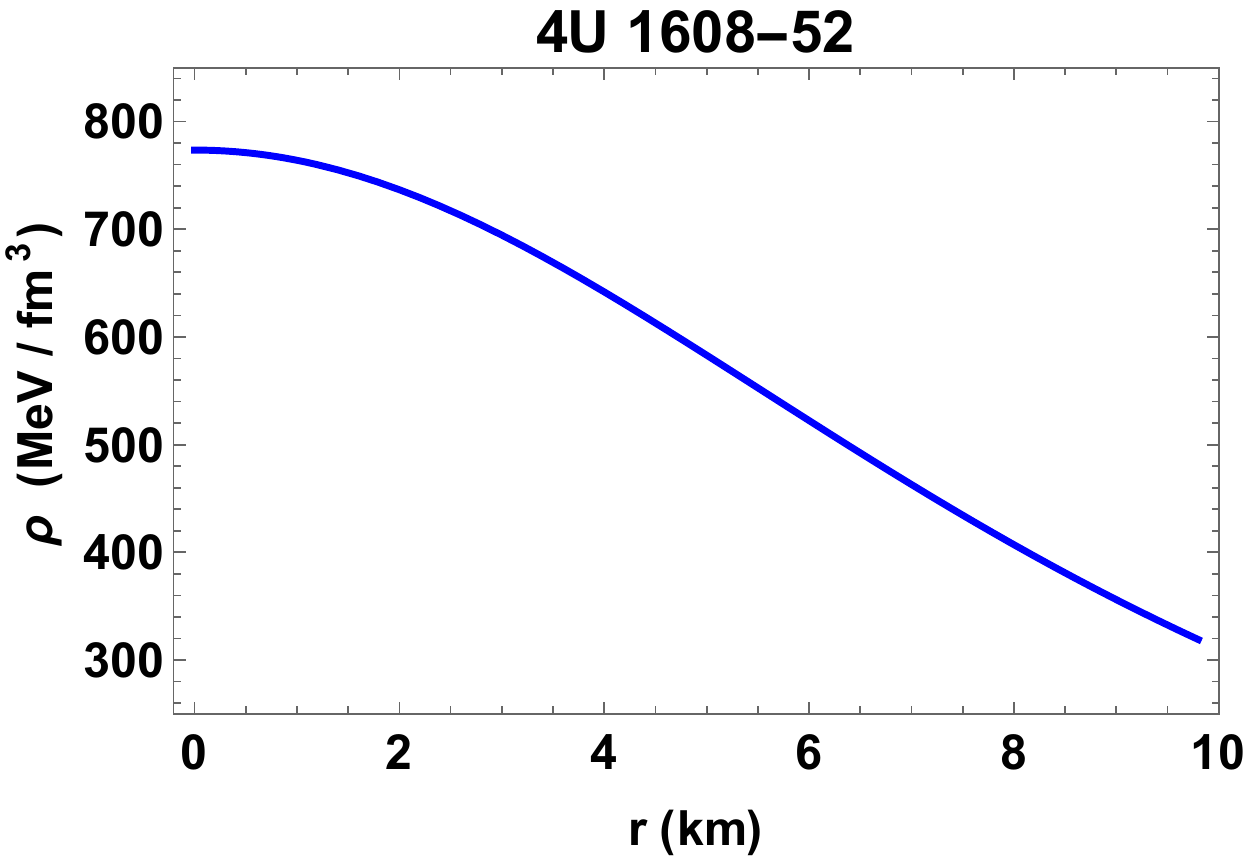}
\includegraphics[width=0.4\columnwidth]{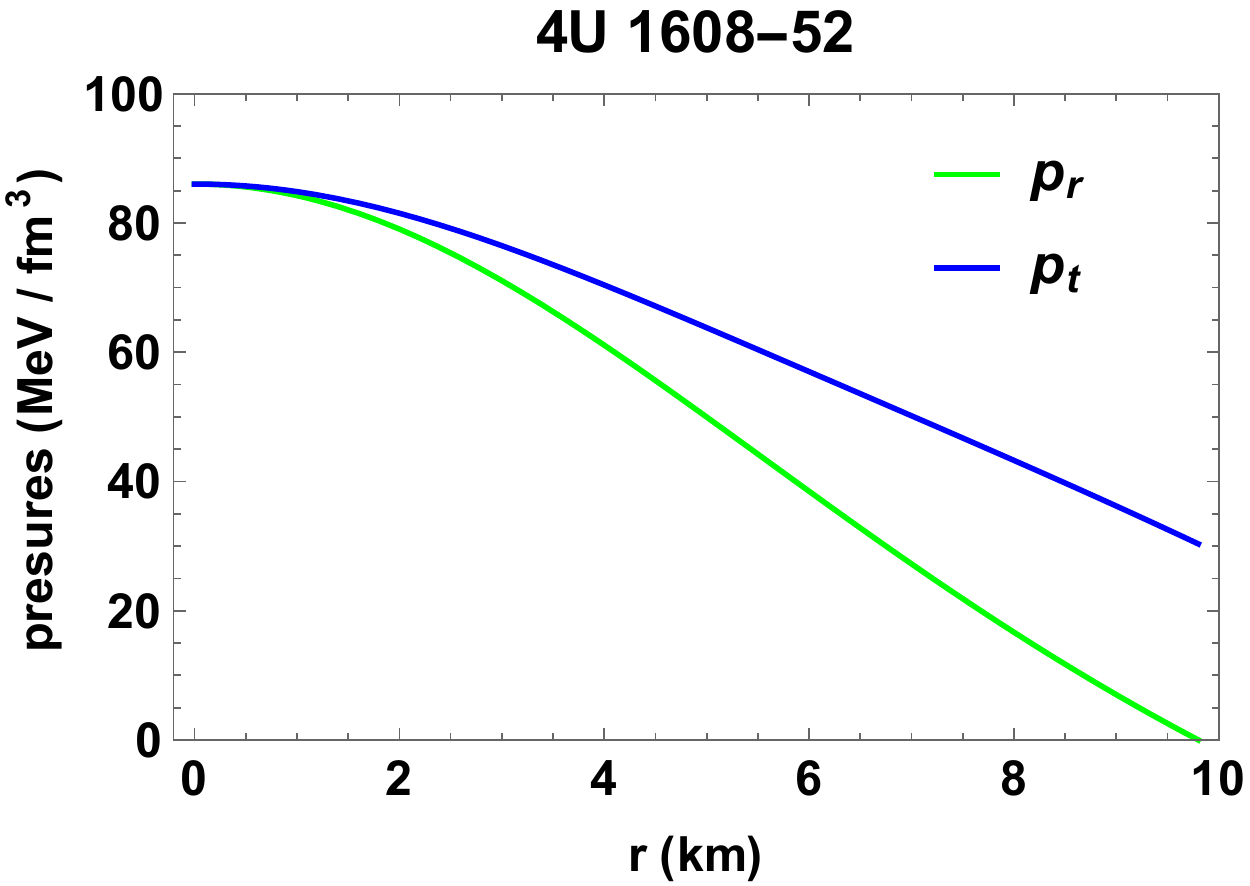}
\caption{Fall-off behaviour of the energy density (left panel) and pressures (right panel).}  \label{fig:Density}
\end{figure}
%%%%%%%%%%%%%%%%%%%%%%%%%%%%%%%%%%%%%%%%%%%%%%%%%%

Now, we would like to take the derivatives of the physical parameters, as follows:
\begin{eqnarray}
\frac{d\rho}{dr}=-\frac{a \text{C}^2 r (2 a+B+2 \delta ) \left(2 a \text{C} r^2+5\right)}{4 \pi  \left(2 a \text{C} r^2+1\right)^3},
\label{drhodr}
\end{eqnarray}

\begin{eqnarray}
\frac{dp_r}{dr}=-\frac{a \alpha  \text{C}^2 r (2 a+B+2 \delta ) \left(2 a \text{C} r^2+5\right)}{4 \pi  \left(2 a \text{C} r^2+1\right)^3},
\label{dprdr}
\end{eqnarray}

Similarly, we can calculate
\begin{align}
\frac{dp_t}{dr}.
\label{dptdr}
\end{align}

%%%%%%%%%%%%%%%%%%%%%%%%%%%%%%%%%%%%%%%%%%%%%%%%%%
\begin{figure}[h]
\centering
\includegraphics[width=0.4\columnwidth]{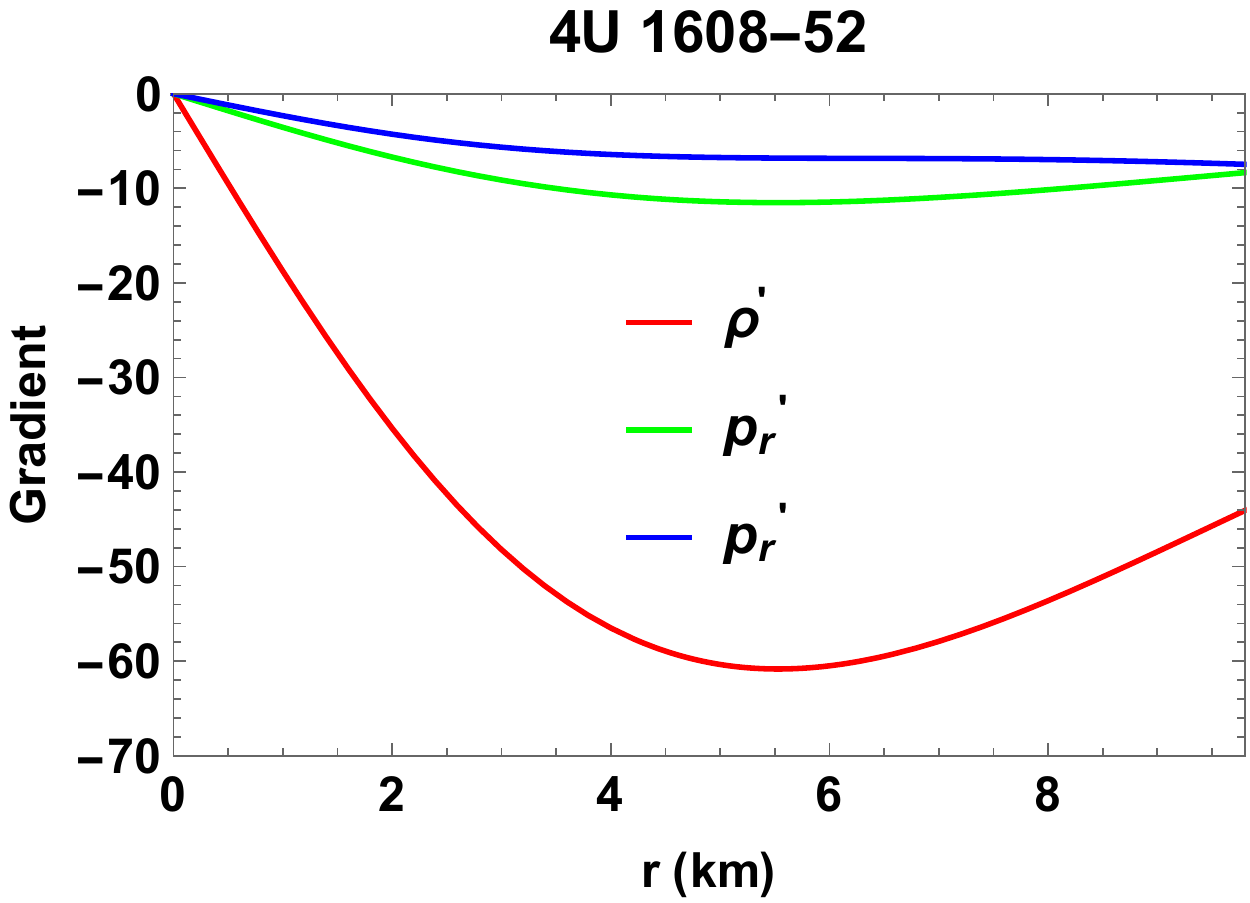}
\includegraphics[width=0.4\columnwidth]{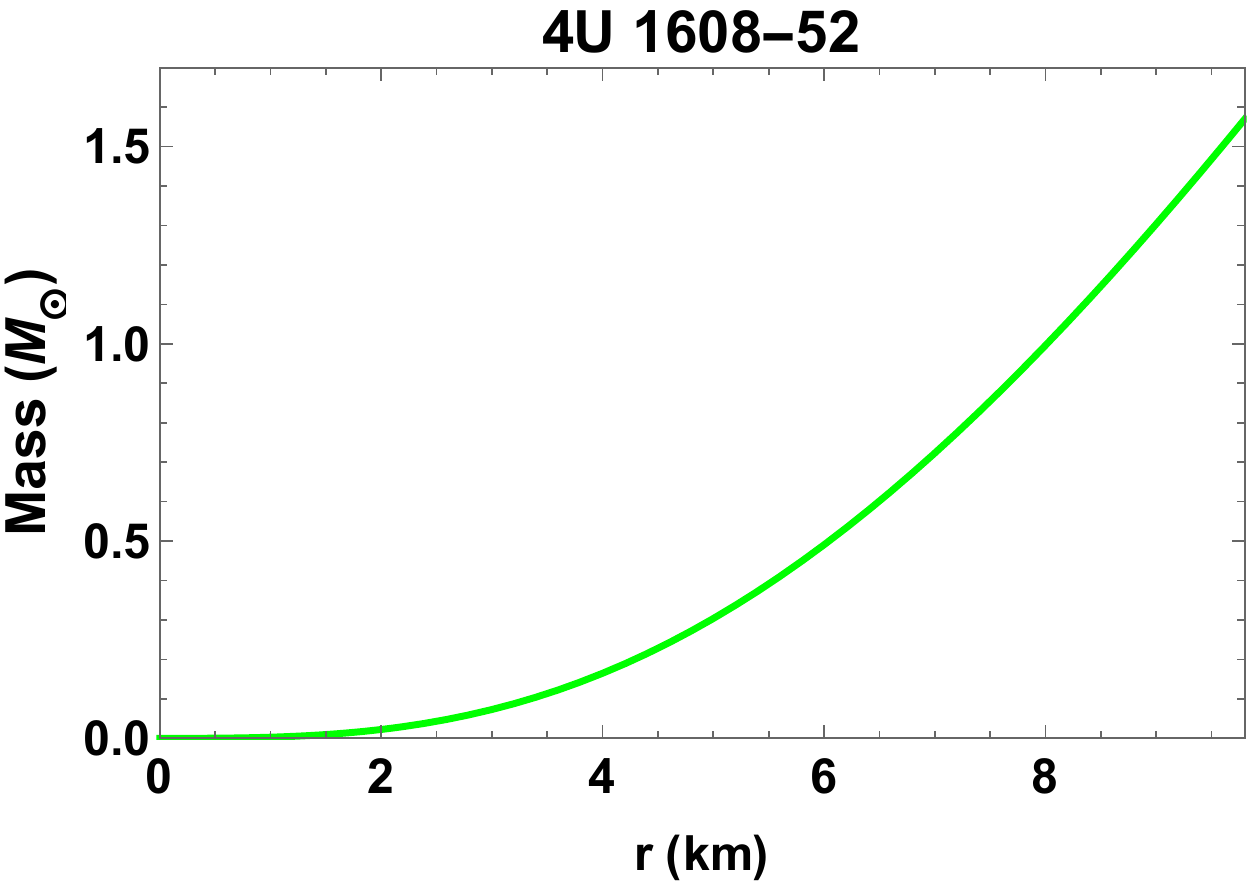}
\caption{Variation of the gradient of the physical parameters (left panel) and mass (right panel) w.r.t. the radial coordinate $r$.} \label{fig:grad}
\end{figure}
%%%%%%%%%%%%%%%%%%%%%%%%%%%%%%%%%%%%%%%%%%%%%%%%%%

With the proper choices of the model parameters within their bound it can be  shown that all the physical parameters, e.g. density, radial pressure and mass show expected behaviors as shown in Figs.~\ref{fig:Density} and~\ref{fig:grad}. 

\end{enumerate}

\section{Physical viability and stability} \label{sec:4}
Let us now check physical viability and analyze stability of the model under the issues as follows.

\subsection{Stability under three different forces}
~~~~~A star remain in static equilibrium under the forces namely, gravitational force ($F_g$), hydrostatics force ($F_h$) and anisotropic force ($F_a$). This condition is formulated mathematically as TOV equation (Tolman-Oppenheimer-Volkoff~\cite{Tolman1939,OV1939}) which is described by the conservation equation given by
\begin{equation}\label{tov1}
\nabla^{\mu}T_{\mu\nu}=0.
\end{equation}

Now using the expression given in (\ref{eq2}) into (\ref{tov1}) one can obtain the following equation:
\begin{equation}\label{tov3}
-\frac{\nu'}{2}(\rho+p_r)+\frac{2}{r}(p_t-p_r)=\frac{dp_r}{dr}.
\end{equation}

Eq.~\eqref{tov3} can be written as
\begin{equation}
F_g+F_h+F_a=0,
\end{equation}
where the expression for $F_g,\,F_h$ and $F_a$ are obtained as:
\begin{align}
F_g &= \dfrac{1}{128 \pi ^2 \left(2 a \text{C} r^2+1\right)^3 \left(\text{C} r^2 \left(a \left(B \text{C} r^2-1\right)+B+\delta \right)-1\right)} \nonumber\\
&\times \left[ \left(\alpha  \text{C} r \left(a \left(\text{C} r^2 \left(a \left(6 B \text{C} r^2+2\right)+7 B+2 \delta \right)+3\right)+3 (B+\delta )\right)\right.\right.\nonumber\\
&\left.+8 \pi  r \left(2 a \text{C} r^2+1\right) \left(\text{C} \left(a r^2 (2 \beta +B \text{C})+a+B+\delta \right)+\beta \right)\right)\nonumber\\
&\times \left(\text{C} \left(2 a^2 \text{C} r^2 \left(\alpha +r^2 (16 \pi  \beta +3 (\alpha +1) B \text{C})+1\right)\right.\right.\nonumber\\
&\left.\left.\left.+a \left(3 (\alpha +1)+r^2 (32 \pi  \beta +(\alpha +1) \text{C} (7 B+2 \delta ))\right)+3 (\alpha +1) (B+\delta )\right)+8 \pi  \beta \right)\right],\\
F_h &= \frac{a \alpha  \text{C}^2 r (2 a+B+2 \delta ) \left(2 a \text{C} r^2+5\right)}{4 \pi  \left(2 a \text{C} r^2+1\right)^3},\\
F_a &= \dfrac{2}{r} \Delta.
\end{align}
where $\Delta = p_t - p_r$ and the expression from $p_r$, $p_t$ are given in Eqs. \eqref{radpres} and \eqref{tangpres}, respectively.

\begin{figure}[h]
\centering
\includegraphics[width=0.4\columnwidth]{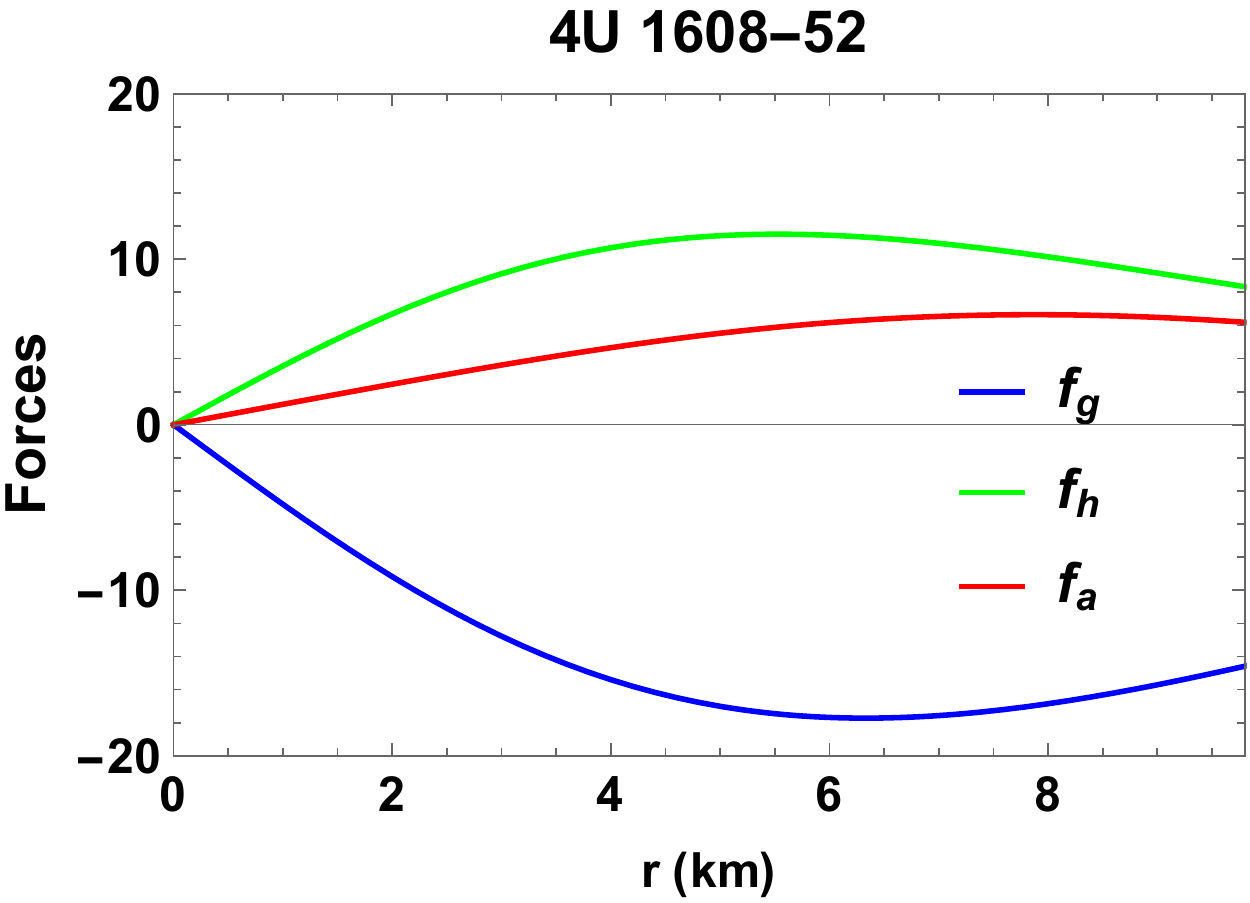}
\includegraphics[width=0.4\columnwidth]{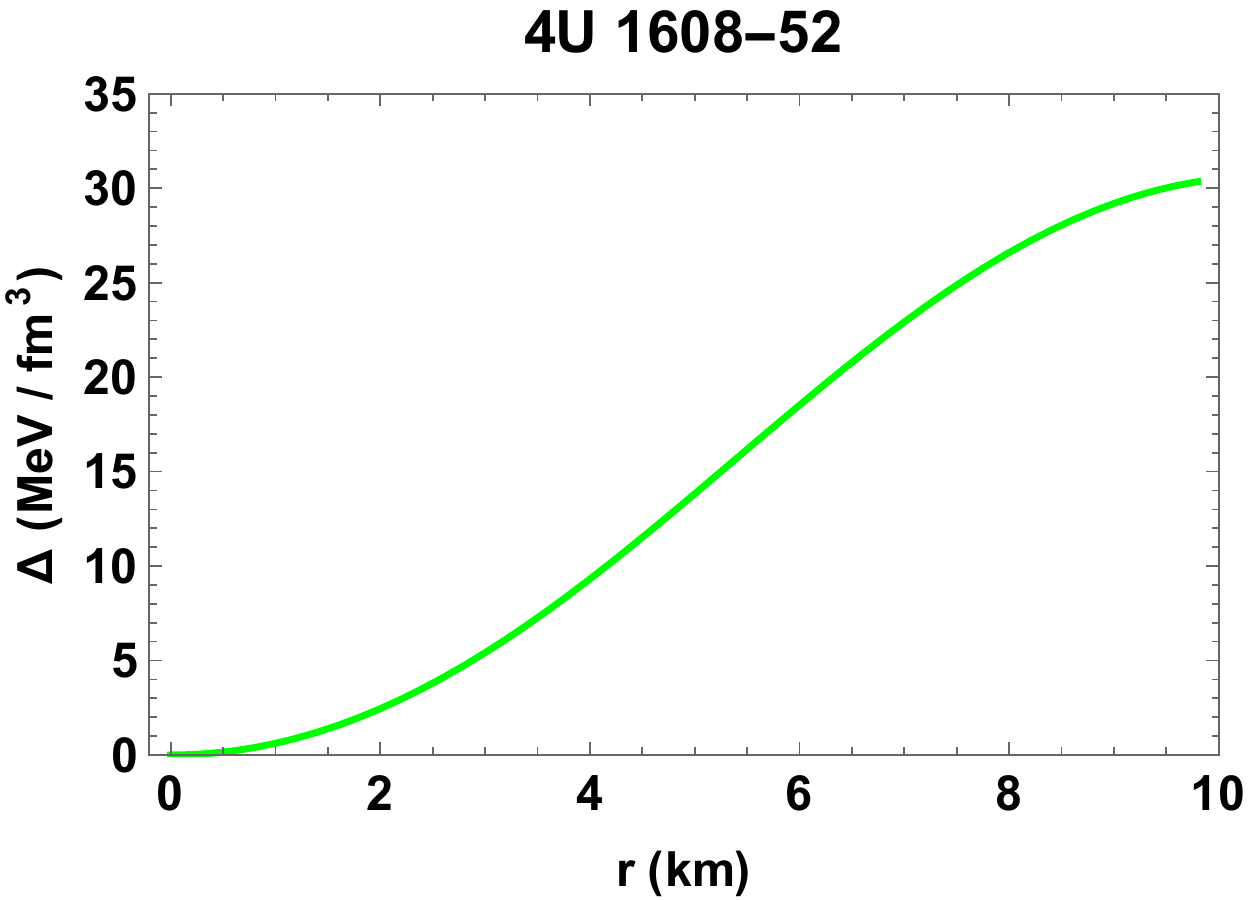}
\caption{Verification of the forces (left panel) and anisotropy parameter (right panel) w.r.t. the radial coordinate $r$.} \label{fig:Strongenergycondition_figure}
\end{figure}

The three different forces are plotted in the left panel of the Fig.~\ref{fig:Strongenergycondition_figure} for the compact star $4U~1608-52$. The figure shows that hydrostatics and anisotropic force are positive and is dominated by the gravitational force which is negative to keep the system in static equilibrium. In the right panel of the figure we have shown feature of the anisotropy parameter.

%%%%%%%%%%%%%%%%%%%%%%%%%%%%%%%%%%%%%%%%%%%%%%%%%%
%%%%%%%%%%%%%%%%%%%%%%%%%%%%%%%%%%%%%%%%%%%%%%%%%%

%%%%%%%%%%%%%%%%%%%%%%%%%%%%%%%%%%%%%%%%%%%%%%%%%%
\begin{figure}[h]
\centering
\includegraphics[width=0.4\columnwidth]{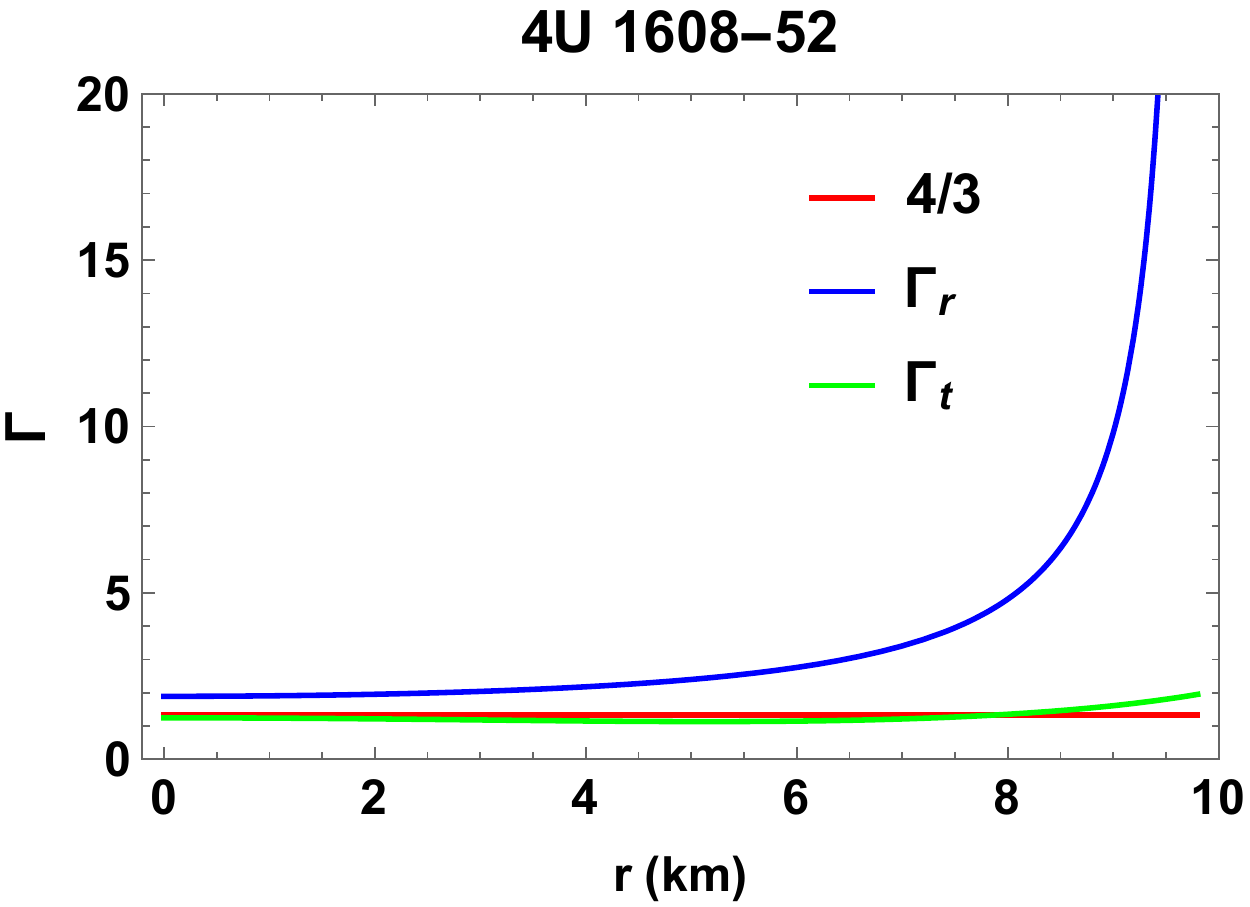}
\caption{Variation of the adiabatic index w.r.t. the radial coordinate $r$.} \label{fig:adiabatic_figure}
\end{figure}
%%%%%%%%%%%%%%%%%%%%%%%%%%%%%%%%%%%%%%%%%%%%%%%%%%

\subsection{Adiabatic index for stability}
~~~~~The adiabatic index which is defined as
\begin{eqnarray}
\Gamma = {\rho(r)+p(r) \over p(r)}{dp(r) \over d\rho(r)},
\end{eqnarray}
is related to the stability of a relativistic anisotropic stellar configuration. 

Any stellar configuration will maintain its stability if adiabatic  index $\Gamma > 4/3$~\cite{Heintzmann}. For our solution, the adiabatic index $\Gamma$ takes the value more than $4/3$ throughout the interior of the compact star, as evident from Fig.~\ref{fig:adiabatic_figure}.

\subsection{Herrera condition for stability}
We also know that for a physically acceptable model, the velocity of the sound (both radial and transverse) should be less than the speed of the light i.e., both $\frac{dp_r}{d\rho},~\frac{dp_t}{d\rho}<1$ which is known as the causality condition.

To examine the stability, we have followed the technique which is known as ``cracking method'' used by Herrera et al.~\cite{LH}. Based on this method Abreau et al.~\cite{Abreu} found that for a compact stellar object, in a stable region we must have 
\begin{eqnarray*}
 v_t^2-v_r^2\left\{
               \begin{array}{ll}
                 <0 ~\text{for}~ 0 \leq r \leq R~ \Rightarrow~& \hbox{potentially stable} \\
                 >0~\text{for}~ 0 \leq r \leq R~\Rightarrow~ &\hbox{unstable}.
               \end{array}
             \right.
\end{eqnarray*}

%%%%%%%%%%%%%%%%%%%%%%%%%%%%%%%%%%%%
\begin{figure}[h]
\centering
\includegraphics[width=0.4\columnwidth]{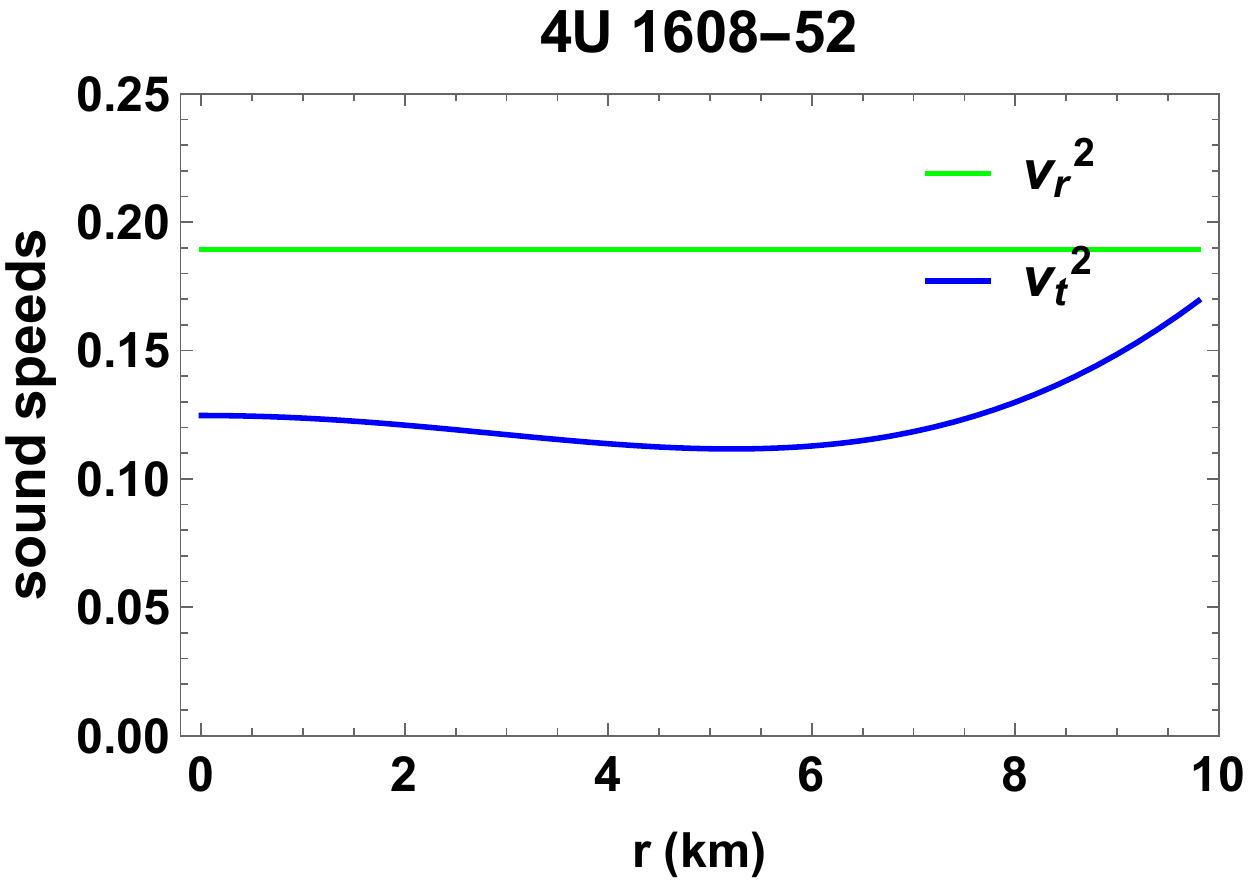}
\includegraphics[width=0.4\columnwidth]{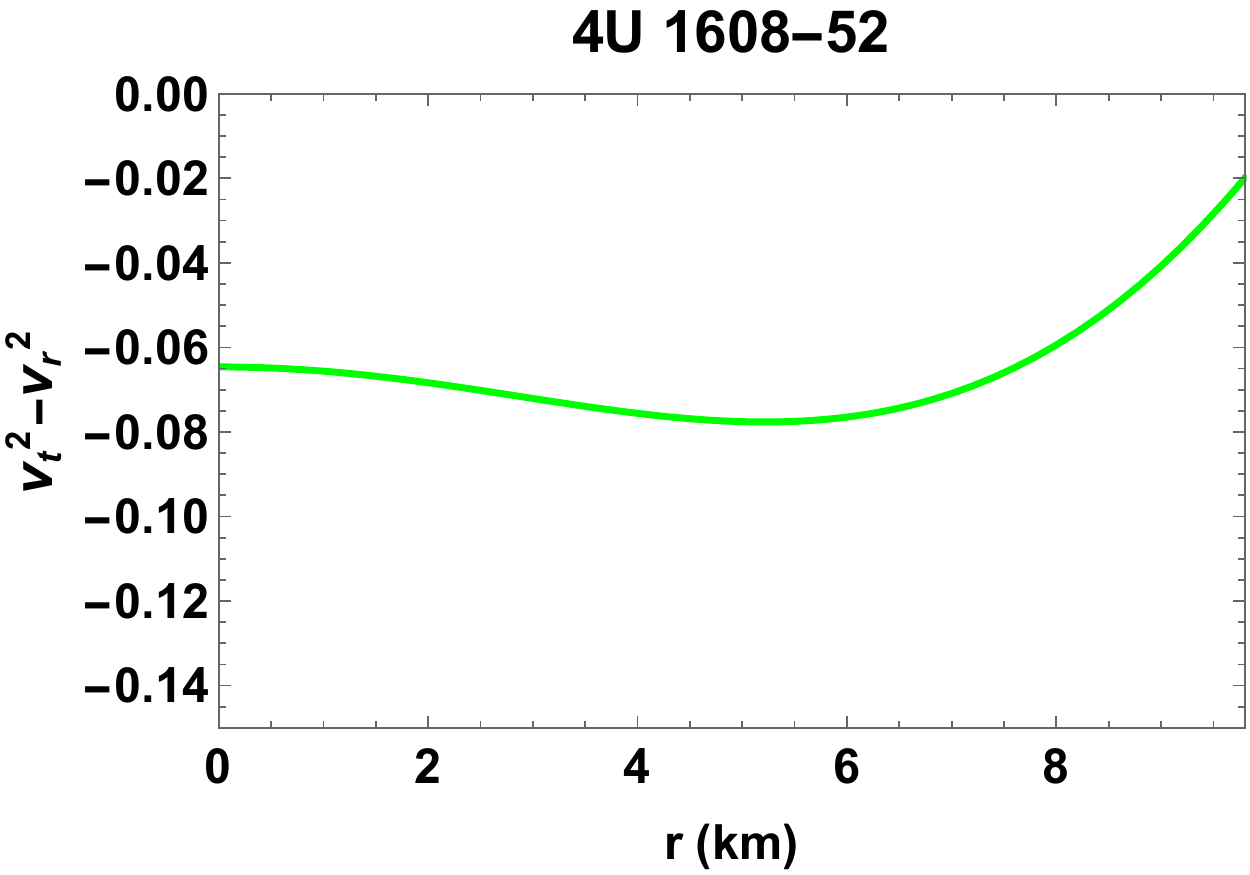}
\caption{Variation of the radial as well as transverse sound speeds (left panel) and the causality condition (right panel) w.r.t. the radial coordinate $r$.}\label{fig:mod_figure}
\end{figure}
%%%%%%%%%%%%%%%%%%%%%%%%%%%%%%%%%%%%

Figure \ref{fig:mod_figure} clearly indicates  that for our assumed set of values the configuration remains stable throughout the star. It is note that at the center of the star, one should get $v_t^2-v_r^2~<0$.

\subsection{Harrison-Zeldovich-Novikov stability condition}
~~~~~Depending on the mass and central density of the star, Harrison et al.~\cite{Harrison} and Zeldovich-Novikov~\cite{ZN} proposed the stability condition for the model of compact star. From their investigation they suggested that for stable configuration $\frac{\partial M}{\partial \rho_c}>0$, where $M,\,\rho_c$ denotes the mass and central density of the compact star.

%%%%%%%%%%%%%%%%%%%%%%%%%%%%%%%%%%%%%%%%%%
\begin{figure}[h]
\centering
\includegraphics[width=0.4\columnwidth]{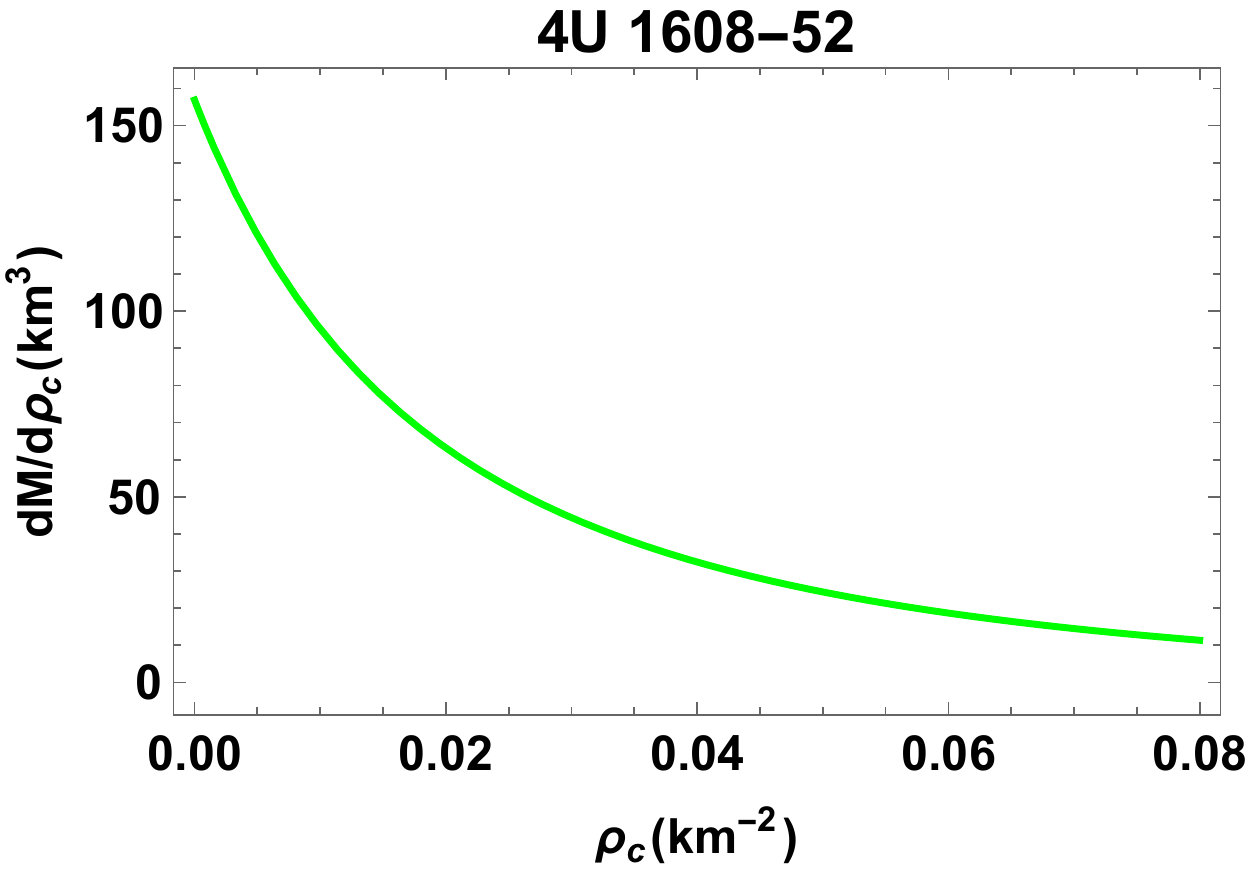}
\caption{Variation of $dM/d\rho_c$ with respect to the central density $\rho_c$.}\label{fig:dmdrho_figure}
\end{figure}
%%%%%%%%%%%%%%%%%%%%%%%%%%%%%%%%%%%%%%%%%%

For our present model
\begin{equation}
\frac{\partial M}{\partial \rho_c}= \frac{R^3 \left(9 a^3+a^2 \left(B \left(2 \rho_c ^2 R^4+6 \rho_c  R^2+27\right)+27 \delta \right)+3 a (B+\delta ) \left(B \left(2 \rho_c R^2+9\right)+9 \delta \right)+9 (B+\delta)^3\right)}{6 (a+B+\delta ) \left(a \left(2 \rho_c  R^2+3\right)+3 (B+\delta )\right)^2}.
\end{equation}

Above expression of $\frac{\partial M}{\partial \rho_c}$ is positive and hence the stability condition is well satisfied which is depicted in Fig.~\ref{fig:dmdrho_figure} with respect to the central density.

\section{Tidal Love number as anisotropic measure} \label{sec:5}
~~~~~In this section we will explore the tidal deformation which is one of the astrophysically observable macroscopic properties and can be used to study the interior of a compact object. Consider an interior metric equation \eqref{metric} of a compact neutron star. Where the expression for $e^{\nu(r)}$ \& $e^{\lambda(r)}$ are given in Eqs. \eqref{etlambda} and \eqref{soln} respectively. Now a small perturbation in the metric due to an infalling objects, i.e. a celestial body, which disturbs the background metric slightly, can be treated as an external tidal field. Because of this external field, the NS will be deformed and hence create a multipolar structure. This scenario can be observed in coalescing binary systems. Mathematically, considering the background metric- $^{(0)}g_{\mu \nu}(x^{\nu})$- metric of a neutron star. The modified metric with small perturbation $h_{\mu \nu}(x^{\nu})$ can be written as
\begin{align}
g_{\mu \nu}\left(x^{\nu}\right)=^{(0)} g_{\mu \nu}\left(x^{\nu}\right)+h_{\mu \nu}\left(x^{\nu}\right), \label{ti1}
\end{align}
where the background geometry of spacetime is 
\begin{align}
^{\left( 0\right)} ds^{2} &=^{\left( 0\right)} g_{\mu \nu }dx^{\mu }dx^{\nu } \nonumber\\ &=-e^{\nu(r) }dt^{2}+e^{\lambda(r) }dr^{2}+r^{2}\left( d\theta ^{2}+\sin ^{2}\theta d\phi ^{2}\right). \label{ti2}
\end{align}

Following the papers~\cite{Biswas19,Regge1957}, for the linearized perturbation $h_{\mu \nu}$, for the sack of simplicity we restrict ourself to static $l=2$,~$m = 0$ even parity perturbation. This restriction is acceptable if the two binary star are sufficiently far away from each other. With these restriction the perturbed metric becomes
\begin{align}
h_{\mu \nu}=\operatorname{diag}\left[H_{0}(r) e^{\nu}, H_{2}(r) e^{\lambda}, r^{2} K(r), r^{2} \sin ^{2} \theta K(r)\right] Y_{2 m}(\theta, \phi). \label{ti3}
\end{align}

As a consequence of external perturbation, the star gets tidally deformed  from its equilibrium position and develops a quadrupole moment $\mathcal{Q}_{ij}$. With linear order approximation, the external tidal field $\mathcal{E}_{ij}$ is related with the quadrupole moment $\mathcal{Q}_{ij}$ as~\cite{Hinderer2008}
\begin{align}
\mathcal{Q}_{ij} = - \Lambda \, \mathcal{E}_{ij}, \label{ti4}
\end{align}
and
\begin{align}
k_{2} = \dfrac{3}{2}\Lambda \, R^{-5}.\label{ti5}
\end{align}
where $\Lambda$ is the tidal deformability of the compact star and it is related to the dimensionless parameter tidal Love number $k_2$.

Based on the following works~\cite{Bowers1974,Doneva2012,Herrera2013,Pretel2020a,Bhar2020,Das2020,Das2021a,Das2021b} and by matching the internal solution with the external solution of the perturbed variable at the surface of the star, one can calculate the final expression for tidal Love number as
 \begin{align}
 		k_2 = [8 (1-2 \mathcal{C})^2 \mathcal{C}^5 (2 \mathcal{C} (\mathit{y}-1)-\mathit{y}+2)]/X, \label{ti17}
\end{align}
 where
\begin{align}
X &= 5(2 \mathcal{C} (\mathcal{C} (2 \mathcal{C} (\mathcal{C} (2 \mathcal{C} (\mathit{y}+1)+3 \mathit{y}-2)-11 \mathit{y}+13)+3 (5 \mathit{y}-8))-3 \mathit{y}+6) \nonumber\\
 		&\left.+3 (1-2 \mathcal{C})^2 (2 \mathcal{C} (\mathit{y}-1)-\mathit{y}+2) \log \left(\frac{1}{\mathcal{C}}-2\right)-3 (1-2 \mathcal{C})^2 (2 \mathcal{C} (\mathit{y}-1)-\mathit{y}+2) \log \left(\frac{1}{\mathcal{C}}\right)\right), \nonumber
\end{align}
with the compactness factor $\mathcal{C} = \dfrac{M}{R}$.
 	 
With the proper choice of different parameters involved in the above equation the numerical value of $k_2$ can be calculated from Eq. $\eqref{ti17}$ as shown in the Table below. 

%%%%%%%%%%%%%%%%%%%%%%%%%%%%%%%%%%%%%%%%%%%%%%%%%%%%%%%%%%%%
\begin{figure}
\centering
\includegraphics[scale=0.4]{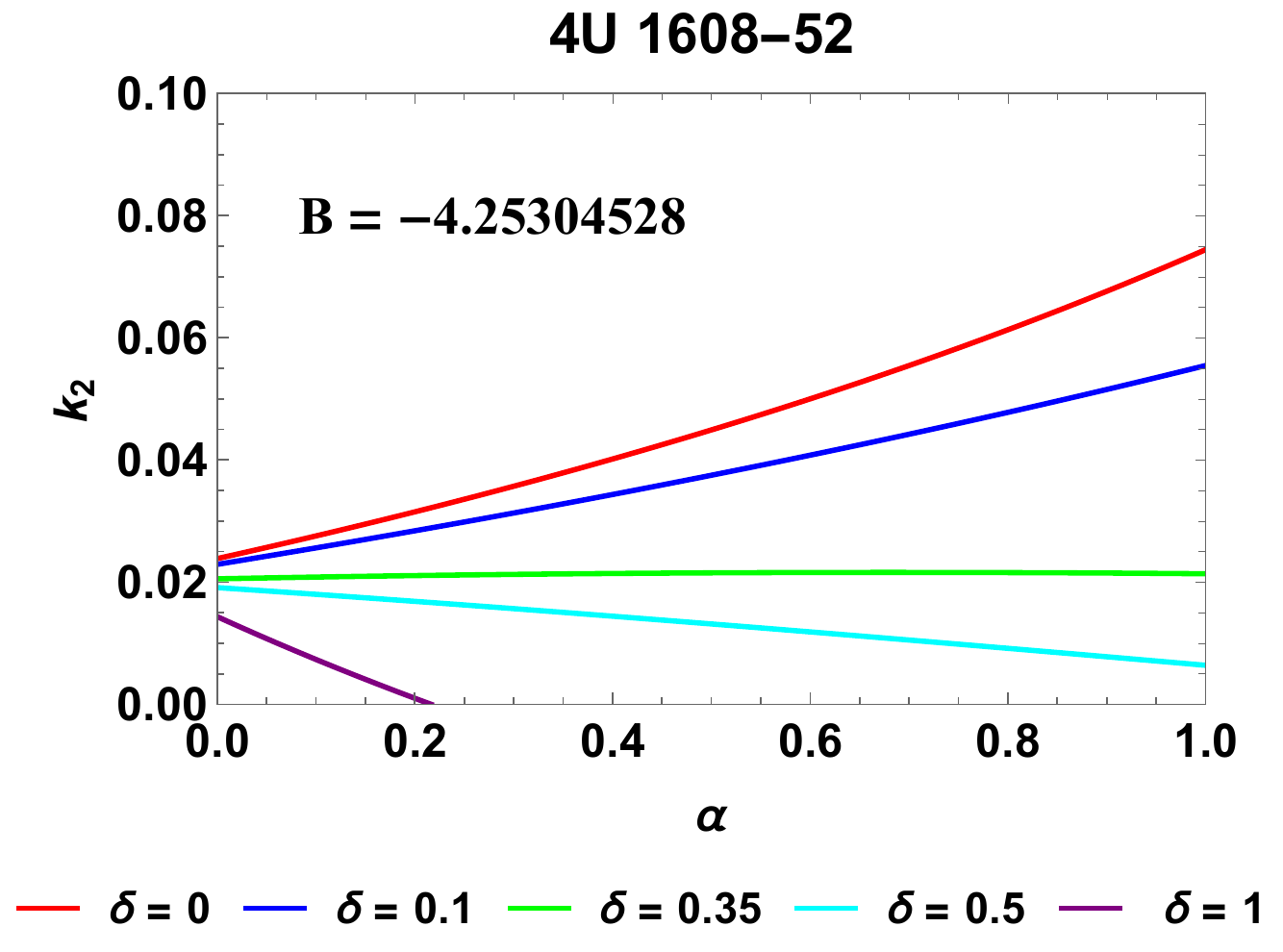}
\includegraphics[scale=0.4]{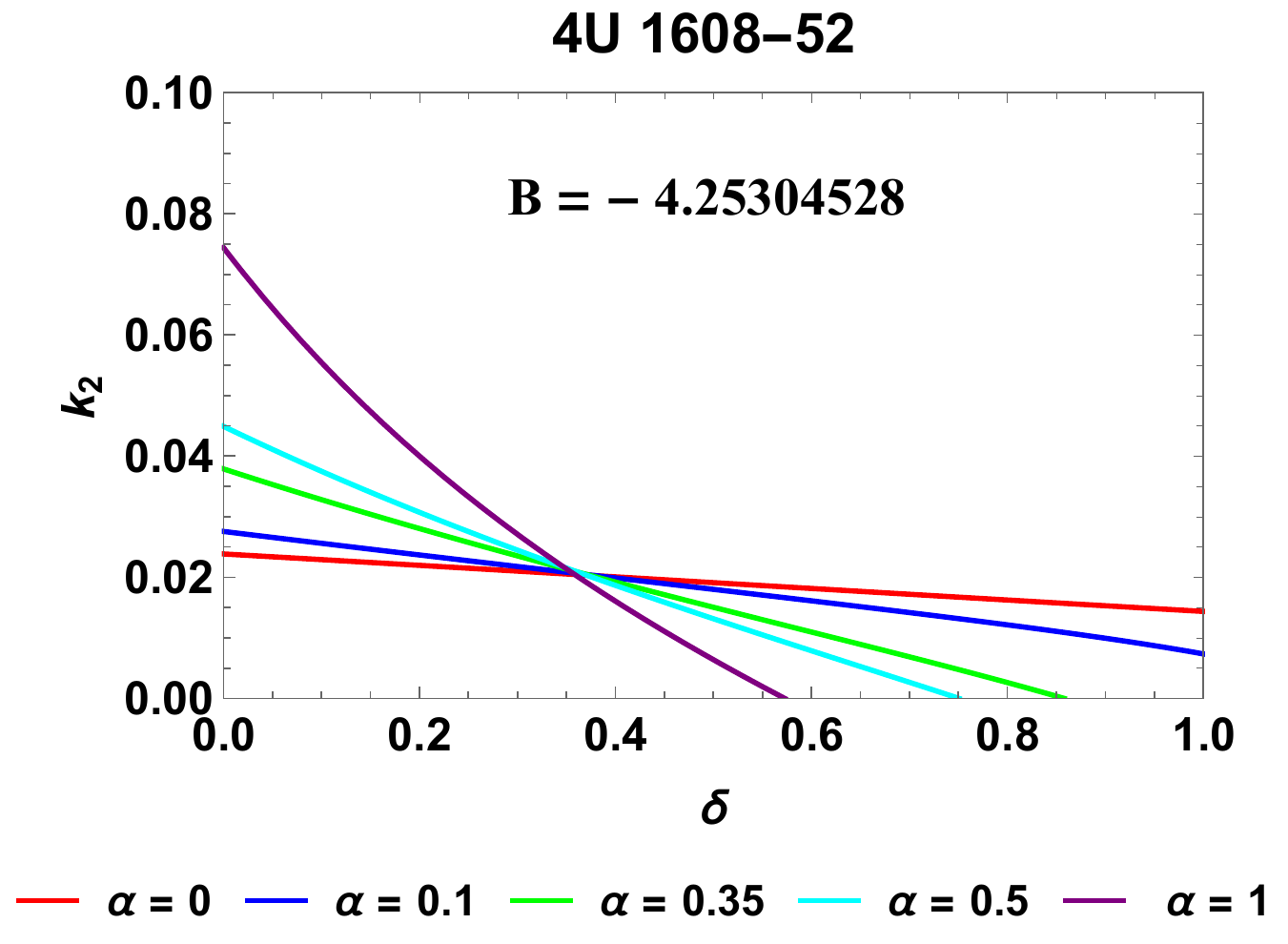}
\includegraphics[scale=0.4]{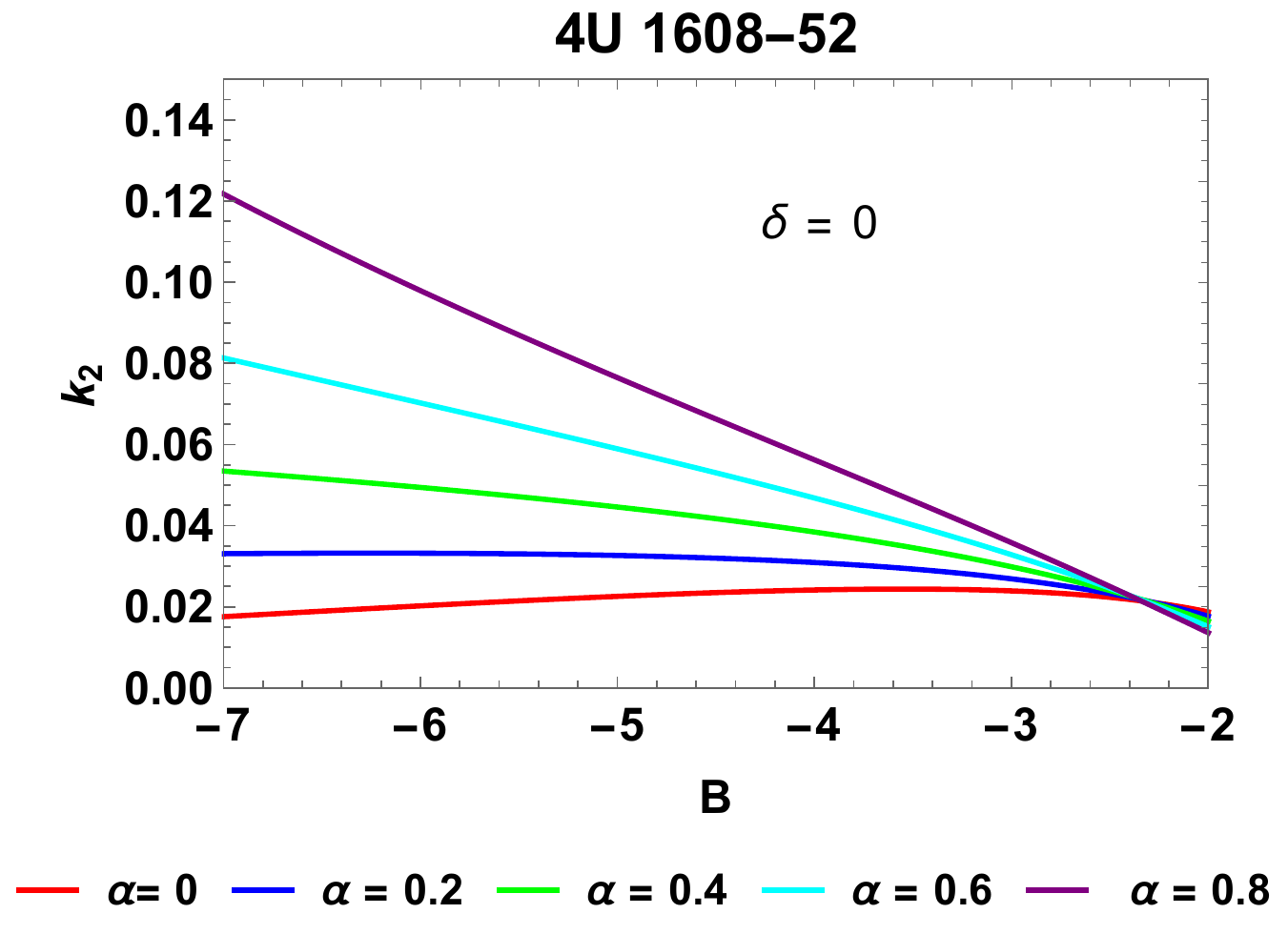}
\includegraphics[scale=0.4]{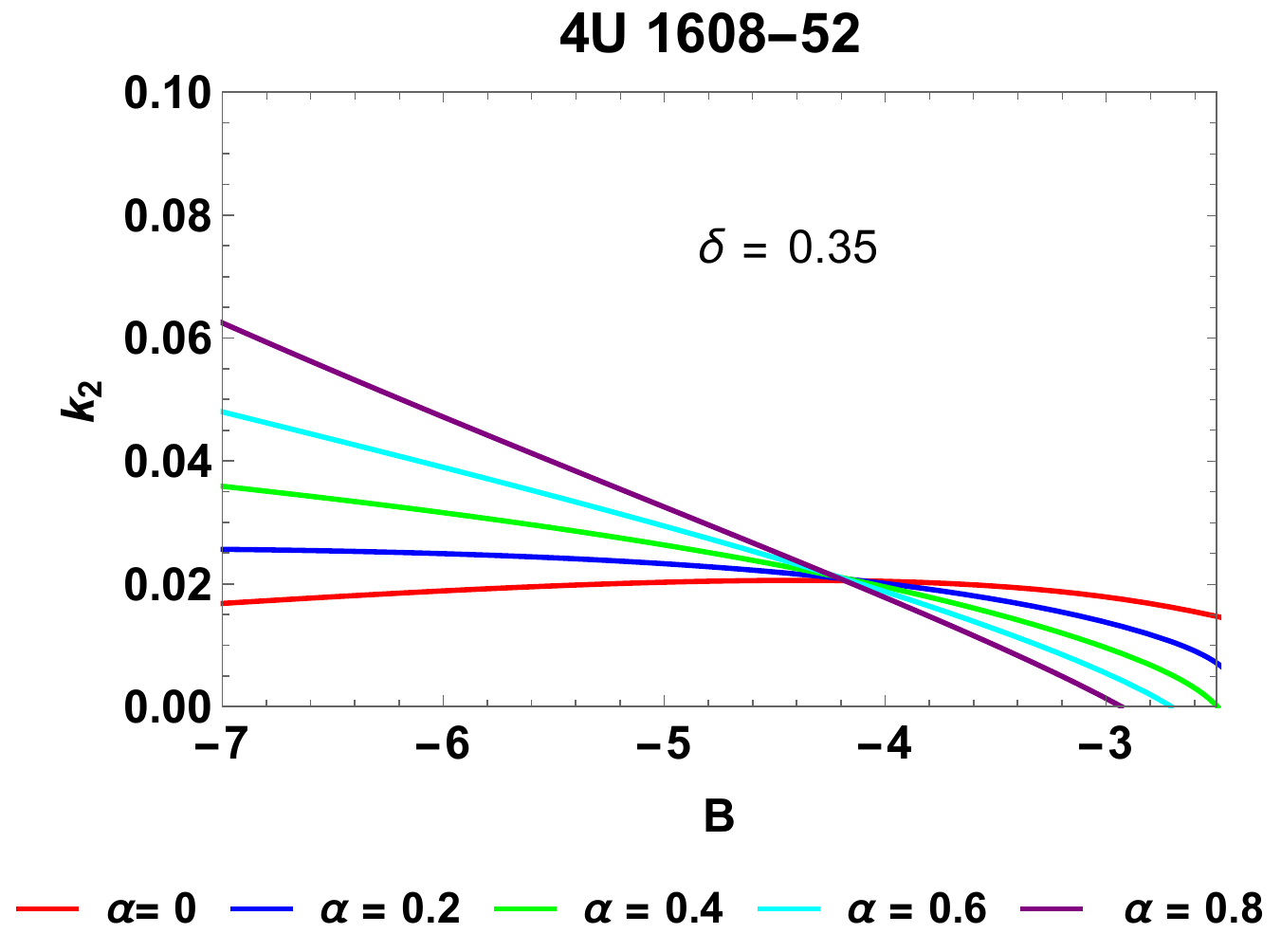}
\includegraphics[scale=0.4]{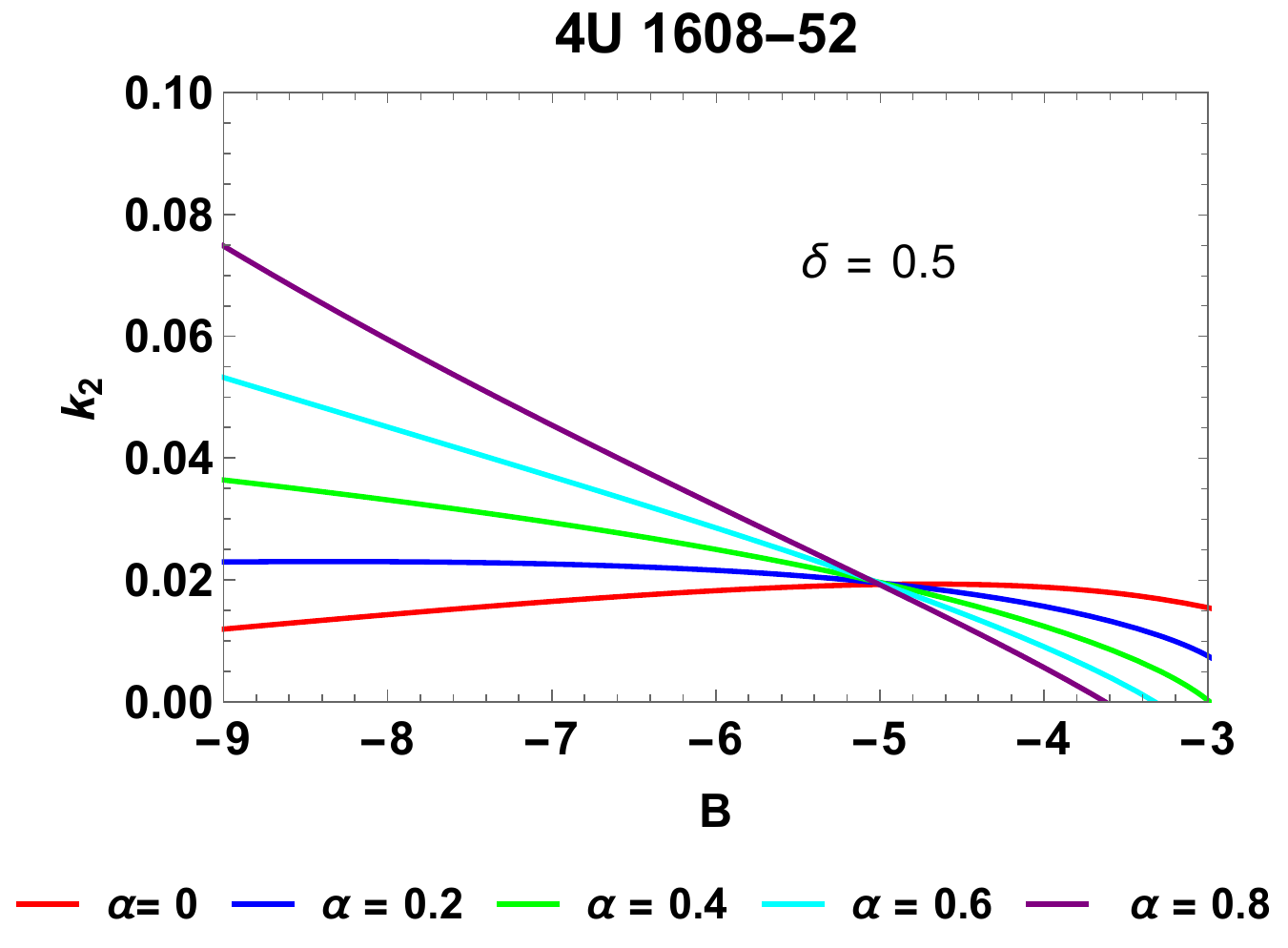}
\includegraphics[scale=0.4]{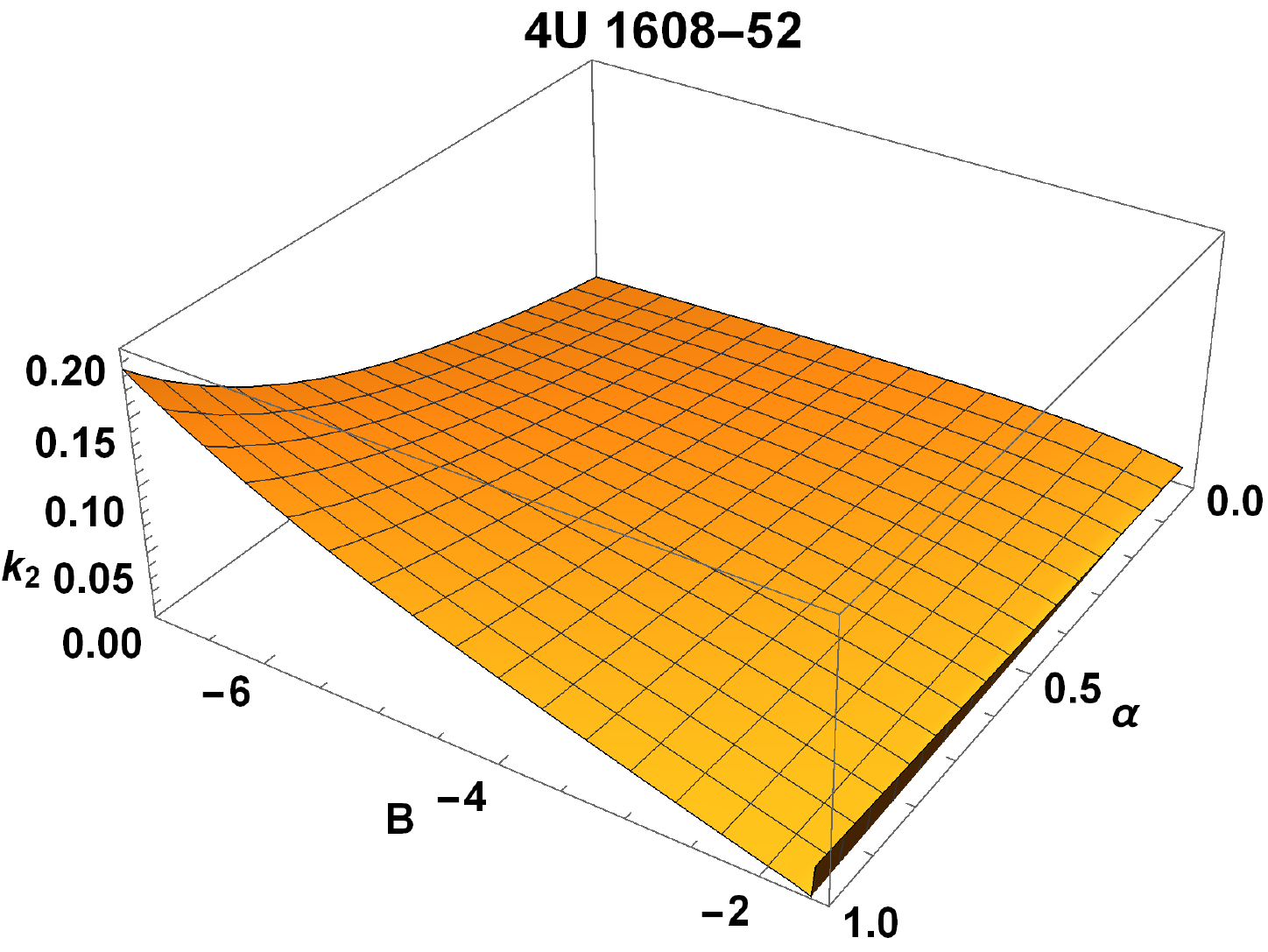}
\includegraphics[scale=0.4]{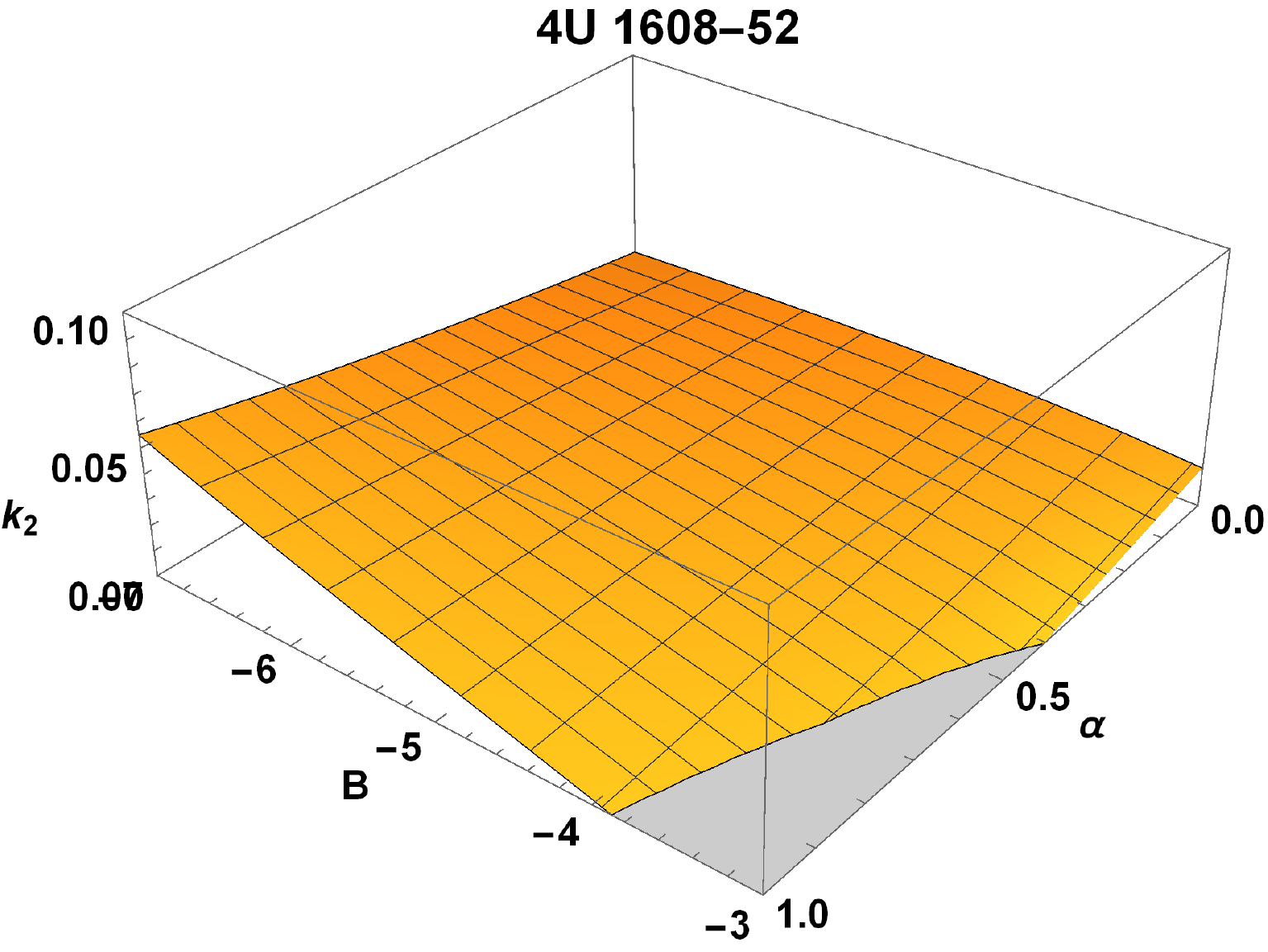}
 \includegraphics[scale=0.35]{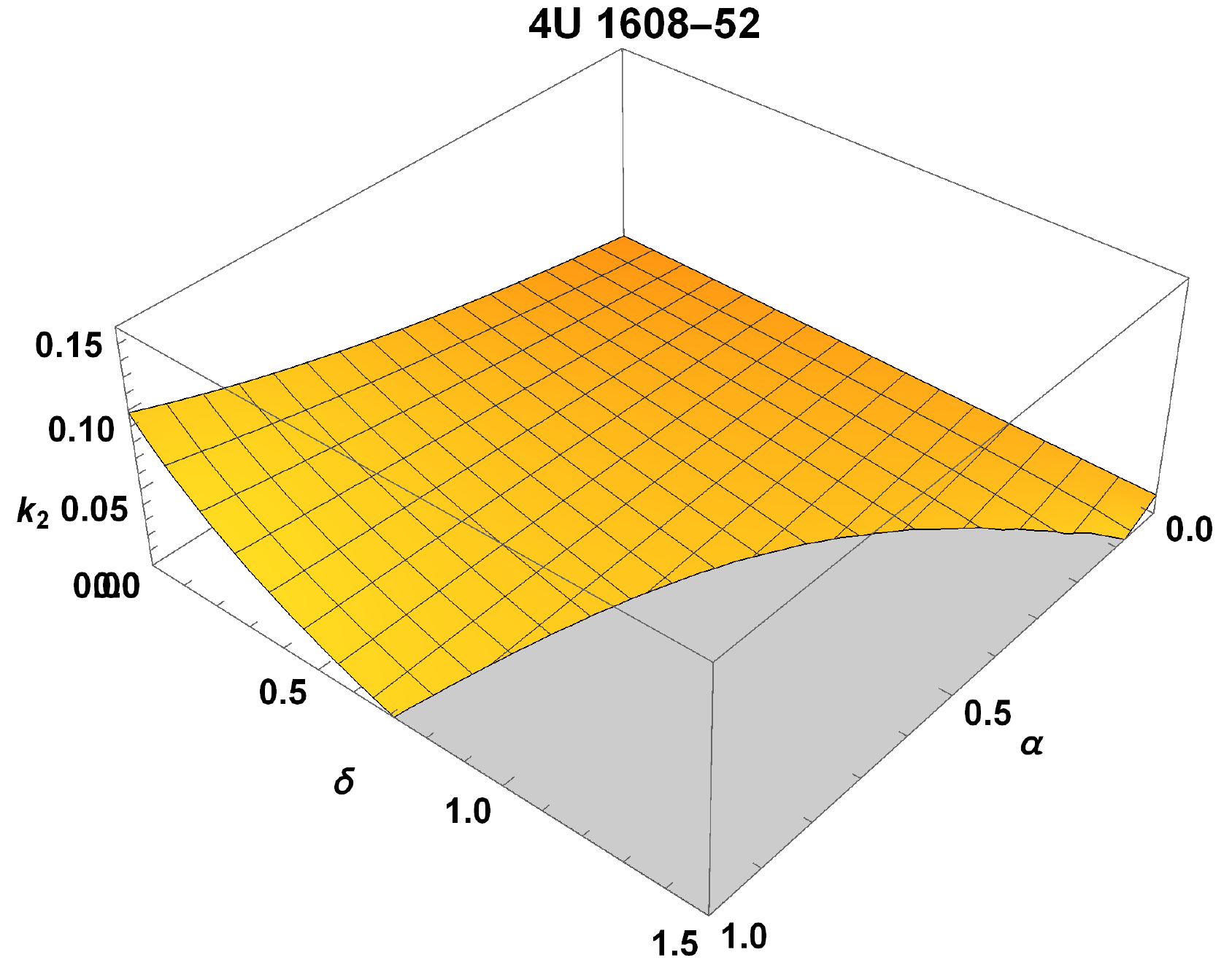}
\caption{Variation of tidal Love number $k_2$ for the compact star $4U~1608-52$ w.r.t. the parameters $\alpha$, $\delta$ and $B$ under the  specific values as indicated in the graphical plots.}\label{fig:k2}
\end{figure}
%%%%%%%%%%%%%%%%%%%%%%%%%%%%%%%%%%%%%%%%%%%%%%%%%%%%%%%%%%%% 	 
 	
%%%%%%%%%%%%%%%%%%%%%%%%%%%%%%%%%%%%%%%%%%%%%%%%%%%%%%%%%%%% 	 	 
\begin{table}[!htp]
\setlength\tabcolsep{5pt} 
\centering 
\caption{ The numerical values of $k_2$ are given for different compact stars under the specific values of $B = -4.253045 $, $\alpha = 0.19$.}
    \begin{tabular*}{\textwidth}{@{\extracolsep{\fill}}lrrrrrrrrl@{}}
    \hline
    Stars & \multicolumn{1}{c}{ M($M_{\odot}$)} & \multicolumn{1}{c}{$R$ (km)}&\multicolumn{1}{c}{$\mathcal{C }= M/R$} & \multicolumn{1}{c}{$k_2$} & \multicolumn{1}{c}{$k_2$} & \multicolumn{1}{c}{$k_2$}\\
& & & & ($\delta = 0$) & ($\delta = 0.35$) & ($\delta = 0.5$)\\ 
\hline 
4U 1608-52   & $1.57^{+0.30}_{- 0.29}$ & $9.8^{+1.8}_{-1.8}$ &0.236301 & 0.031090 & 0.021051 &  0.016970  \\

4U 1820-30 & $1.46^{+0.21}_{-0.21}$ & $11.1^{+1.8}_{-1.8}$ & 0.194009& 0.038194 &0.023408 &  0.017369\\
      
4U 1724-207 & $1.81^{+0.25}_{-0.37}$ & $12.2^{+1.4}_{-1.4}$ &0.218832& 0.034082 & 0.022211 & 0.017382  \\

KS 1731-260 & $1.61^{+0.35}_{-0.37}$ & $10^{+2.2}_{-2.2}$ &0.237475& 0.030897 & 0.020964 & 0.016933 \\ 
\hline \label{tab1}
\end{tabular*}
\end{table}
%%%%%%%%%%%%%%%%%%%%%%%%%%%%%%%%%%%%%%%%%%%%%%%%%%%%%%%%%%%% 	 

In the above Table, the observational values of masses and radii of the compact stars have been employed from the reference of Roupas and Nashed~\cite{Roupas2020} such that the numerical values of $k_2$ are given for different compact stars under the specific values of $B = -4.253045 $, $\alpha = 0.19$.

\section{Discussion} \label{sec:6}
~~~~~In this paper, we have obtained a class of interior solutions to the Einstein field equations for an anisotropic matter distribution obeying a linear EOS. The solution seems interesting being regular and well-behaved  and hence could describe a relativistic compact star. 

To show that the solution can be used as a viable model for compact observed sources, we consider the pulsar $4U~1608-52$ whose mass and radius are estimated to be $M = 1.57_{-0.29}^{+0.30}~M_\odot $ and $R = 9.8_{-1.8}^{+1.8}~$km, respectively~\cite{Roupas2020}. For the given mass and radius, we have determined the values of the constants $C=-0.0070827$, $\beta=-0.002$ and $A=0.00956915$ for arbitrarily chosen values of $\alpha=0.19$, $B=-4.25304528$ and $\delta=1.2$.  For physical acceptability of our model, using the values of the constants and plugging the values of $G$ and $c$, we have tried to figure out the behaviour of the physically relevant quantities graphically within the stellar interior. Therefore, based on the graphical plots, which usually depict basic features of a given model, we would like to mention some salient features of our presented model as follows:

(i) Left panel of Fig.~\ref{fig:metrices} shows that the metric potentials are positive within the stellar interior as per the requirement whereas in the right panel we have shown the regular feature of the potentials on the boundary. 

(ii) Verification of the energy condition w.r.t. the radial coordinate $r$ has been done in Fig. \ref{fig:energycondition_figure} which is satisfactory as far as physical criteria are concerned. 

(iii) Fig. \ref{fig:Density} show variations of the energy density $\rho$, radial pressure $p_r$ and tangential pressure $ p_t $, respectively in the left and right panels. The pressures are radially decreasing outwards from its maximum value at the centre and in case of radial pressure it drops to zero at the boundary as is expected but the tangential pressure remains non-zero at the boundary. Obviously, all the quantities decrease monotonically from the centre towards the boundary. On the other hand, variation of the gradient of these physical parameters  (left panel) as well as the mass (right panel) are shown in Fig. \ref{fig:grad}. Note that the mass function is regular at the center.

(iv) Verification of the forces (left panel) and anisotropy parameter (right panel) w.r.t. the radial coordinate $r$ have been depicted with their expected unique features in Fig.~\ref{fig:Strongenergycondition_figure}. It can be noted that the outwardly acting combined anisotropic and hydrostatic forces balance the inwardly acting strong gravitational force. Variation of anisotropy indicates that it is zero at the centre as usual and is maximum at the surface. 

(v) In Figs. \ref{fig:adiabatic_figure} and \ref{fig:mod_figure} we have demonstrated features of the adiabatic index, the radial as well as transverse sound speeds (left panel) and the causality condition (right panel) w.r.t. the radial coordinate $r$. Similarly, variation of $dM/d\rho_c$ with respect to the central density $\rho_c$ has been shown in Fig. \ref{fig:dmdrho_figure}.

(vi) In the Fig. \ref{fig:k2}, the parameter $k_2$ is plotted against $\alpha$ for the compact star $4U~1608-52$ under the specific value of $B= -4.25304528$ and $\delta$. It is evident from the figure that for a particular value of $B$, for range $0\leq \delta < 0.3581$, tidal Love number $k_2$ increases  with increasing $\alpha$ and  for $\delta > 0.3581$, $k_2$ decreases with increasing $\alpha$. However, for $\delta = 0.3581$, $k_2$ remains approximately  constant with increasing $\alpha$ values. This also reflects in other panel, i.e. the curves with different $\alpha$ values intersect at the same point, in this case specifically at $\delta = 0.35$. The same thing reflects in the Fig. \ref{fig:k2}, i.e. for a particular choice of $\delta = 0.50$, the curves with different $\alpha$ values intersect at the point $B = -4.25304528$.
From the different plots it is therefore clear that with increasing $\delta$ values, the intersection point of the curves with different $\alpha$ values shift towards more negative $B$ values.
 	
In the lower panels we have presented some interesting 3D plots to understand the pattern of Love number and hence in turn role of anisotropy which is responsible for the tidal effect. 

In the provided Table \ref{tab1}, the numerical values of $k_2$ are given for different compact objects. It is transparent from the table that with increasing compactness $\mathcal{C}$ of NS, the tidal Love number decreases.
 	 
Finally, it is very interesting to mention here that the range of $k_2$ for the compact star $4U~1608-52$ resembles with the numerical values of $k_2$ as obtained in the paper~\cite{Yazadjiev2018}. As in our previous work~\cite{Das2021b} here also we would like to mention that (i) in the presented compact star the gravitational tidal effect is solely responsible for the pressure anisotropy and (2) an observational $k_2$ would imply a maximum possible core mass and metallicity which may indicate interesting internal structure of the compact stars~\cite{Kramm2011}. \\

\section*{acknowledgments}
SR, SD and KC gratefully acknowledge support from the Inter-University Centre for Astronomy and Astrophysics (IUCAA), Pune, India under its Visiting Research Associateship Programme.

\section*{Appendix A}

\label{appendix}
\begin{align}
\frac{dp_t}{d\rho}&=\dfrac{-1}{4
   a \text{C}^2 \left(2 a \text{C} r^2+1\right) \left(2 a \text{C} r^2+5\right) (2 a+B+2 \delta ) \varrho_{1}^2} \times\nonumber\\
   &\left[8 a^6 \text{C}^6 \varrho_{2} r^8-4 a^5 \text{C}^5 \left(B^3 \text{C}^3 \left(3 \alpha ^2+2 \alpha -1\right) r^6+2 B^2 \text{C}^2 \varrho_{3} r^4+B \text{C} \varrho_{4} r^2\right.\right.\nonumber\\
 &  \left.+2 \left(4 \text{C} \beta ^2 \delta  r^6-16 \beta ^2 r^4+(\text{C} \delta -2 \beta ) r^2+2 \alpha  \left(\text{C}
   \delta  r^2+2\right)+\alpha ^2 \left(\text{C} \delta  r^2+5\right)+3\right)\right) r^6\nonumber\\
   &+2 a^4 \text{C}^4 \left(2 B^3 \text{C}^3 \left(7 \alpha ^2+16 \alpha +5\right)
   r^6-32 \text{C} \beta ^2 \delta  r^6+96 \beta ^2 r^4+4 \text{C}^2 \delta ^2 r^4+16 \text{C} \beta  \delta  r^4\right.\nonumber\\
   &+B^2 \text{C}^2 \varrho_{5} r^4+24 \beta  r^2-12 \text{C} \delta  r^2+4
   B \text{C} \varrho_{6}
   r^2+\alpha ^2 \left(4 \text{C}^2 \delta ^2 r^4-20 \text{C} \delta  r^2-27\right)\nonumber\\
   &\left.+2 \alpha  \varrho_{7}-7\right) r^4+a^3 \text{C}^3 \left(B^3 \text{C}^3 \left(85 \alpha ^2+110 \alpha +33\right) r^6+2 B^2 \text{C}^2 \varrho_{8} r^4+B \text{C} \varrho_{9} r^2\right.\nonumber\\
   &+2 \left(4 \text{C}
   \delta  \left(-6 \beta ^2+2 \text{C} \delta  \beta +\text{C}^2 \delta ^2\right) r^6+4 \left(18 \beta ^2+8 \text{C} \delta  \beta +3 \text{C}^2 \delta ^2\right) r^4+(26
   \beta -9 \text{C} \delta ) r^2\right.\nonumber\\
   &\left.\left.+\alpha ^2 \left(4 \text{C}^3 \delta ^3 r^6+20 \text{C}^2 \delta ^2 r^4-33 \text{C} \delta  r^2-6\right)+2 \alpha  \varrho_{10}+2\right)\right) r^2+\beta ^2+3
   \text{C}^2 \delta ^2\nonumber\\
   &+9 \text{C}^2 \alpha ^2 \delta ^2+12 \text{C}^2 \alpha  \delta ^2+3 B^2 \text{C}^2 \left(3 \alpha ^2+4 \alpha +1\right)+4 \text{C} \beta  \delta +6
   \text{C} \alpha  \beta  \delta +2 B \text{C} \varrho_{11}\nonumber\\
   &+a^2 \text{C}^2 \left(24
   \text{C}^3 \delta ^3 r^6+16 \text{C}^2 \beta  \delta ^2 r^6+2 B^3 \text{C}^3 \left(35 \alpha ^2+14 \alpha +11\right) r^6-16 \text{C} \beta ^2 \delta  r^6\right.\nonumber\\
   &+58 \beta ^2
   r^4+6 \text{C}^2 \delta ^2 r^4+56 \text{C} \beta  \delta  r^4+B^2 \text{C}^2 \varrho_{12} r^4+24 \beta  r^2+8 \text{C} \delta  r^2+2 B \text{C} \varrho_{13} r^2\nonumber\\
   &\left.+\alpha ^2 \left(40
   \text{C}^3 \delta ^3 r^6+30 \text{C}^2 \delta ^2 r^4-24 \text{C} \delta  r^2+9\right)+4 \alpha  \varrho_{14}+3\right)\nonumber\\
  & +a \text{C} \left(B^3 \text{C}^3 \left(21 \alpha ^2-14 \alpha +5\right)
   r^4+2 B^2 \text{C}^2 \varrho_{15} r^2+B \text{C} \varrho_{16}+2 \left(\left(6 r^2-\text{C} r^4 \delta \right) \beta ^2\right.\right.\nonumber\\
   &+\varrho_{17}\beta +\text{C} \delta  \left(5 \text{C}^2 \delta ^2 r^4+2 \text{C} \delta  r^2-2 \alpha 
   \left(7 \text{C}^2 \delta ^2 r^4-38 \text{C} \delta  r^2+14\right)\right.\nonumber\\
   &\left.\left.\left.\left.+3 \alpha ^2 \left(7 \text{C}^2 \delta ^2 r^4-2 \text{C} \delta  r^2+3\right)+3\right)\right)\right)\right],
\end{align}
where\\
$\varrho_{1} = B \text{C} r^2+a \text{C} \left(B \text{C} r^2-1\right)
   r^2+\text{C} \delta  r^2-1$,\\
$\varrho_{2} =4 \beta ^2 r^4+B^2 \text{C}^2 \left(15 \alpha ^2+22 \alpha +7\right) r^4+2 B \text{C} \left(6 \beta  r^2+\alpha ^2+\alpha  \left(8 \beta r^2+2\right)+1\right) r^2\\
  -\alpha ^2-2 \alpha -1$,\\
$\varrho_{3} = \left((2 \beta
   -\text{C} \delta ) r^2+\alpha ^2 \left(3 \text{C} r^2 \delta -62\right)+2 \alpha  \left(r^2 (2 \beta +\text{C} \delta )-45\right)-24\right)$,\\
$\varrho_{4} = \left(4 \beta 
   (\beta +2 \text{C} \delta ) r^4-8 (11 \beta +\text{C} \delta ) r^2-\alpha ^2 \left(8 \text{C} \delta  r^2+17\right)\right.\\
  \left. +2 \alpha  \left(8 \text{C} \beta  \delta  r^4-8 (8
   \beta +\text{C} \delta ) r^2-9\right)-17\right)$,\\
$\varrho_{5} = \left(-8 (\beta -3 \text{C} \delta )
   r^2+\alpha ^2 \left(32 \text{C} \delta  r^2+373\right)+\alpha  \left(490-8 r^2 (2 \beta -9 \text{C} \delta )\right)+141\right)$,\\
$\varrho_{6} = \left(-4 \beta ^2-4 \text{C} \delta  \beta +2 \text{C}^2 \delta ^2\right) r^4+(62 \beta +17 \text{C} \delta ) r^2+\alpha ^2 \left(2 \text{C}^2 \delta
   ^2 r^4+25 \text{C} \delta  r^2+15\right)+\alpha  \left(4 \text{C} \delta  (\text{C} \delta -2 \beta ) r^4+(92 \beta +42 \text{C} \delta ) r^2+31\right)+10$,\\
$\varrho_{7} = \left(4 \text{C} \delta  (2 \beta +\text{C} \delta ) r^4+8 (\beta -3 \text{C}
   \delta ) r^2-21\right)$,\\
$\varrho_{8} = \left((2 \beta +49 \text{C} \delta) r^2+3 \alpha ^2 \left(39 \text{C} \delta  r^2+95\right)+2 \alpha  \left((2 \beta +87 \text{C} \delta ) r^2+167\right)+105\right)$,\\
$\varrho_{9} = 4 \left(-6 \beta
   ^2+4 \text{C} \delta  \beta +17 \text{C}^2 \delta ^2\right) r^4+40 (9 \beta +4 \text{C} \delta ) r^2+\alpha ^2 \left(132 \text{C}^2 \delta ^2 r^4+320 \text{C} \delta r^2+81\right)+\alpha  \left(8 \text{C} \delta  (4 \beta +33 \text{C} \delta ) r^4+48 (11 \beta +8 \text{C} \delta ) r^2+378\right)+49$,\\
$\varrho_{10} = 4 \text{C}^2
   \delta ^2 (2 \beta +\text{C} \delta ) r^6+8 \text{C} \delta  (2 \beta +\text{C} \delta ) r^4+(14 \beta -\text{C} \delta ) r^2-34$,\\
$\varrho_{11} = \left((3 \alpha +2) \beta +3 \text{C} \left(3 \alpha ^2+4 \alpha +1\right) \delta \right)$,\\
$\varrho_{12} =(8 \beta +78 \text{C} \delta ) r^2+\alpha ^2 \left(230 \text{C} \delta r^2+231\right)+2 \alpha  \left((8 \beta +66 \text{C} \delta ) r^2+181\right)+83$,\\
$\varrho_{13} = -4 \left(\beta ^2-3 \text{C} \delta  \beta -10 \text{C}^2 \delta ^2\right) r^4+(75 \beta +37 \text{C} \delta ) r^2+\alpha ^2 \left(100 \text{C}^2 \delta ^2 r^4+93 \text{C} \delta 
   r^2+18\right)+2 \alpha  \left(2 \text{C} \delta  (6 \beta +25 \text{C} \delta ) r^4+(54 \beta +101 \text{C} \delta ) r^2+39\right)+12$,\\
$\varrho_{14} = 8 \text{C}^2 \delta ^2 (\beta +3 \text{C} \delta ) r^6+\text{C}
   \delta  (16 \beta +13 \text{C} \delta ) r^4+4 (2 \beta +\text{C} \delta ) r^2-17$,\\
$\varrho_{15} =(\beta +10 \text{C} \delta ) r^2+6 \alpha ^2 \left(7 \text{C} \delta  r^2+4\right)+2 \alpha  \left((\beta -14 \text{C} \delta )
   r^2+37\right)+10$,\\
$\varrho_{16} = \left(-\beta ^2+6 \text{C} \delta  \beta +25 \text{C}^2 \delta ^2\right) r^4+12 (3 \beta +2 \text{C} \delta ) r^2+3 \alpha
   ^2 \left(35 \text{C}^2 \delta ^2 r^4+12 \text{C} \delta  r^2+6\right)+2 \alpha  \left(\text{C} \delta  (6 \beta -35 \text{C} \delta ) r^4+2 (13 \beta +75 \text{C} \delta ) r^2-8\right)+6$,\\
$\varrho_{17} = 2 \left(\text{C}^2 \delta ^2 r^4+6 \text{C} \delta  r^2+1\right)+\alpha \left(4 \text{C}^2 \delta ^2 r^4+16 \text{C} \delta  r^2+3\right).$\\

\end{document}